\documentclass[utf8]{FrontiersinHarvard} % for articles in journals using the Harvard Referencing Style (Author-Date), for Frontiers Reference Styles by Journal: https://zendesk.frontiersin.org/hc/en-us/articles/360017860337-Frontiers-Reference-Styles-by-Journal
\usepackage{url,hyperref,lineno,microtype,subcaption}
\usepackage[onehalfspacing]{setspace}
\usepackage{amsmath}
\def\chandra{\textit{Chandra}}
\def\xmm{\textit{XMM-Newton}}
\def\nustar{\textit{NuSTAR}}
\def\soxs{\textsc{soxs}}
\def\sixte{\textsc{sixte}}

\def\keyFont{\fontsize{8}{11}\helveticabold }
\def\firstAuthorLast{Pfeifle {et~al.}} %use et al only if is more than 1 author
\def\Authors{Ryan W. Pfeifle\,$^{1,2,\ddagger{},*}$, Peter G. Boorman\,$^{3}$, Kimberly A. Weaver\,$^{1}$, Johannes Buchner\,$^{4}$, Francesca Civano\,$^{1}$, Kristin Madsen\,$^{1}$, Daniel Stern\,$^{5}$, N\'uria Torres-Alb\`a$^{6}$, Emanuele Nardini$^{7}$, Claudio Ricci\,$^{8,9,10}$, Stefano Marchesi\,$^{6}$, D. R. Ballantyne\,$^{11}$, Dominic Sicilian\,$^{12}$, Chien-Ting Chen\,$^{13,14}$, Elias Kammoun\,$^{15,16,7}$, Ryan C. Hickox$^{17}$, Javier A. Garc\'ia\,$^{1,3}$, Labani Mallick\,$^{18,19,3,\dagger}$}
\def\Address{ \footnotesize $^{1}$X-ray Astrophysics Laboratory, NASA Goddard Space Flight Center,Code 662, Greenbelt, MD, USA \\
$^{2}$Oak Ridge Associated Universities, NASA NPP Program, Oak Ridge, TN, USA \\
$^\ddagger{}$ NASA Postdoctoral Program Fellow \\
$^{3}$Cahill Center for Astrophysics, California Institute of Technology, 1216 East California Boulevard, Pasadena, CA 91125, USA \\
$^{4}$Max Planck Institute for Extraterrestrial Physics, Giessenbachstrasse, 85748 Garching, Germany \\
$^{5}$Jet Propulsion Laboratory, California Institute of Technology, Pasadena, CA, 91109, USA \\
$^{6}$Department of Physics and Astronomy, Clemson University, Kinard Lab of Physics, Clemson, SC 29634, USA\\
$^{8}$Instituto de Estudios Astrof\'isicos, Facultad de Ingenier\'ia y Ciencias, Universidad Diego Portales, Av. Ej\'ercito Libertador 441, Santiago, Chile\\
$^{9}$Kavli Institute for Astronomy and Astrophysics, Peking University, Beijing 100871, China\\
$^{10}$George Mason University, Department of Physics and Astronomy, MS3F3, 4400 University Drive, Fairfax, VA 22030, USA\\
$^{11}$ Center for Relativistic Astrophysics, School of Physics, George Institute of Technology, 837 State Street, Atlanta, GA, 30332-0430, USA \\
$^{12}$Department of Physics, University of Miami, Coral Gables, FL 33146 \\
$^{13}$Science and Technology Institute, Universities Space Research Association, Huntsville, AL 35805, USA\\
$^{14}$Astrophysics Office, NASA Marshall Space Flight Center, ST12, Huntsville, AL 35812, USA\\
$^{15}$IRAP, Universit\'{e} de Toulouse, CNRS, UPS, CNES, 9, Avenue du Colonel Roche, BP 44346, F-31028, Toulouse Cedex 4, France \\
$^{16}$Dipartimento di Matematica e Fisica, Universit\`{a} Roma Tre, via della Vasca Navale 84, I-00146 Rome, Italy\\
$^{17}$Department of Physics and Astronomy, Dartmouth College, Hanover, New Hampshire 03755, USA \\
$^{18}$Department of Physics \& Astronomy, University of Manitoba, Winnipeg, Manitoba R3T 2N2, Canada\\
$^{19}$Canadian Institute for Theoretical Astrophysics, University of Toronto, 60 St George Street, Toronto, Ontario M5S 3H8, Canada \\
$^{\dagger}$ CITA National Fellow
}
\def\corrAuthor{Ryan W. Pfeifle}

\def\corrEmail{ryan.w.pfeifle@nasa.gov}

\begin{document}
\onecolumn
\firstpage{1}

\title {The High Energy X-ray Probe (HEX-P): The Future of Hard X-ray Dual AGN Science} 

\author[\firstAuthorLast ]{\Authors} %This field will be automatically populated
\address{} %This field will be automatically populated
\correspondance{} %This field will be automatically populated

\extraAuth{}% If there are more than 1 corresponding author, comment this line and uncomment the next one.
%\extraAuth{corresponding Author2 \\ Laboratory X2, Institute X2, Department X2, Organization X2, Street X2, City X2 , State XX2 (only USA, Canada and Australia), Zip Code2, X2 Country X2, email2@uni2.edu}

\maketitle

\begin{abstract}
%%% Leave the Abstract empty if your article does not require one, please see the Summary Table for full details.

A fundamental goal of modern-day astrophysics is to understand the connection between supermassive black hole (SMBH) growth and galaxy evolution. Merging galaxies offer one of the most dramatic channels for galaxy evolution known, capable of driving inflows of gas into galactic nuclei, potentially fueling both star formation and central SMBH activity. Dual active galactic nuclei (dual AGNs) in late-stage mergers with nuclear pair separations $<10$\,kpc are thus ideal candidates to study SMBH growth along the merger sequence since they coincide with the most transformative period for galaxies. However, dual AGNs can be extremely difficult to confirm and study. Hard X-ray ($>10$\,keV) studies offer a relatively contamination-free tool for probing the dense obscuring environments predicted to surround the majority of dual AGN in late-stage mergers. To date, only a handful of the brightest and closest systems have been studied at these energies due to the demanding instrumental requirements involved. We demonstrate the unique capabilities of \textit{HEX-P} to spatially resolve the soft and - for the first time - hard X-ray counterparts of closely-separated ($\sim2''-5''$) dual AGNs in the local Universe. By incorporating state-of-the-art physical torus models, we reproduce realistic broadband X-ray spectra expected for deeply embedded accreting SMBHs. Hard X-ray spatially resolved observations of dual AGNs – accessible only to \textit{HEX-P} – will hence transform our understanding of dual AGN in the nearby Universe.

\tiny
 \keyFont{ \section{Keywords:} Dual AGN, Galaxy Merger, X-ray Astronomy, Active Galactic Nucleus, Galaxy Interaction} %All article types: you may provide up to 8 keywords; at least 5 are mandatory.
\end{abstract}

\section{Introduction}
\label{sec:intro}

It has long been known that galaxy interactions and mergers give rise to gravitational torques that can transform the structure and composition of the constituent galaxies \citep[e.g.,][]{toomre1972}. These torques can funnel vast quantities of gas into the central nuclei of the galaxies \citep[e.g.,][]{barnes1991,barnes1996}, creating reservoirs that can then fuel star formation as well as the central supermassive black holes (SMBHs), which ignite as active galactic nuclei (AGNs) \citep{hopkins2006,hopkins2008}. A natural product of galaxy mergers are dual AGNs, that is, synchronized SMBH growth in both galaxies \citep{vanwassenhove2012}. Simulations of galaxy mergers predict that the growth of the central SMBHs is better synchronized in the latest stages of the merger sequence, when the central nuclei are separated by $\lesssim10$\,kpc \citep{vanwassenhove2012,capelo2015,blecha2018}, though dual AGNs can be found across the full merger sequence, from separations below $1$\,kpc \citep[e.g.,][]{komossa2003,koss2023} and up $\sim50$\,kpc or more \citep{liu2011,koss2012,derosa2018,derosa2023}, and the synchronized activation of a dual AGN is a strong function of progenitor mass ratios, morphological types, gas masses, and radiative feedback \citep[e.g.,][]{callegari2009,callegari2011,vanwassenhove2012,capelo2015,blecha2018,li2021}. Dual AGNs represent a potentially important phase of SMBH growth: they are predicted to be heavily obscured across the merger sequence \citep{blecha2018}  and in late-stage mergers they are expected to coincide with the most obscured and expeditious growth of the two SMBHs \citep{blecha2018,capelo2015}. These predictions find support in observations showing that, relative to mass- and redshift-matched control galaxies, the AGN fraction in galaxy mergers increases with decreasing nuclear pair separation \citep{ellison2011,satyapal2014,weston2017}, where the increase in AGN fraction is most dramatic when selecting AGNs based on mid-IR colors \citep{satyapal2014,weston2017}. A large fraction of known dual AGNs show evidence for heavy obscuration \citep[see Section~5.2 in ][]{pfeifle2023c} which is consistent with theoretical predictions, but such evidence is admittedly still relatively circumstantial and may depend upon selection method; statistically complete studies of dual AGNs are still needed. Nonetheless, with additional evidence of dramatic AGN feedback in known dual AGNs \citep{muller-sanchez2018}, dual AGNs in late-stage mergers may represent one of the most transformative phases of the entire merger sequence for both the SMBHs and their hosts. Dual AGNs also represent the observational forerunner to gravitationally-bound binary SMBHs, the inspiral and coalescence of which will emit the most titanic of gravitational wave signals (GW) and will be accessible to future space-based GW facilities \citep[e.g., LISA,][which will be sensitive to intermediate-mass black hole binaries]{amaro-seoane2023}. Despite this presumed importance, dual AGNs have remained relatively elusive, and a variety of selection strategies have been pursued to uncover them: optical spectroscopic emission line ratios \cite{liu2011}, double-peaked emission lines \cite[e.g.,][]{zhou2004,wang2009,liu2010,smith2010,comerford2011,comerford2012,ge2012}, infrared colors \citep{imanishi2014,satyapal2017,ellison2017,pfeifle2019a,pfeifle2019b,imanishi2020,barrows2023}, hard X-ray selection \citep{koss2012}, and most recently varstrometry \citep{hwang2020,shen2021,schwartzman2023} and Gaia multi-source strategies \citep{shen2019,mannucci2022,ciurlo2023}. Serendipitous \citep[e.g,][]{komossa2003,guainazzi2005,piconcelli2010} and follow-up X-ray observations \citep{bianchi2008,comerford2011,koss2012,liu2013,comerford2015,satyapal2017,pfeifle2019a,pfeifle2019b,hou2019,gross2019,hou2020,hou2023} are a common method by which dual AGNs are confirmed\footnote{Radio imaging is also an effective method of selecting \citep{fu2015a} as well as confirming or rejecting dual AGN candidates \citep[e.g.,][]{fu2011,gabanyi2014,frey2012,gabanyi2016,fu2015b,mullersanchez2015,rubinur2019}.}, particularly in cases where dual AGNs cannot be selected through common optical diagnostics \cite[e.g., NGC 6240,][]{komossa2003}. These investigations have revealed a prevalence of high column densities in the known population of dual AGNs \citep[e.g.,][]{komossa2003,bianchi2008,piconcelli2010,ptak2015,nardini2017,derosa2018,pfeifle2019a,pfeifle2019b,derosa2023}; see discussion in \citet[][]{pfeifle2023c}.

While \chandra{} and \xmm{} have been used effectively in the past to identify, confirm, and/or characterize dual AGN systems, their soft X-ray bandpasses ($0.5-10$\,keV) bias them against detecting and properly characterizing the X-ray spectral properties and the circumnuclear obscurers in heavily obscured dual AGNs with column densities $N_{\rm{H}}>5\times10^{23}-10^{24}$ cm$^{-2}$ cm$^{-2}$, as predicted \citep{hopkins2008,capelo2015,capelo2017,blecha2018} and observed \citep[e.g.,][]{komossa2003,ballo2004,bianchi2008,mazzarella2012,koss2016,derosa2018,iwasawa2018,pfeifle2019b,guainazzi2021,iwasawa2020,derosa2023}, particularly in late-stage mergers \citep[e.g.,][]{ricci2017MNRAS,ricci2021}. In fact, this is a particular concern in LIRGs and ULIRGs; though the statistics are small, a higher fraction of dual AGNs were identified among the lower IR luminosity GOALS sample \citep{torres-alba2018} than the brighter IR luminosity sample \citep{iwasawa2011}, which can possibly be interpreted as the dual AGNs in the high luminosity sample being too obscured for \textit{XMM-Newton} or \textit{Chandra} to detect and  too close together to be resolved by \textit{NuSTAR} \citep{torres-alba2018}. Hard X-ray bandpasses ($>10$\,keV), on the other hand, provide access to harder X-ray features, such as Compton reflection, that are vital for providing accurate constraints on the X-ray properties of heavily obscured AGNs. \textit{NuSTAR}, with its sensitivity over the 3-78\,keV bandpass, has detected and studied the hard X-ray dual AGN counterparts (when present) in Arp 299 \citep[though only one of the dual AGNs was detected,][but see Section~\ref{sec:arp299} below]{ptak2015}, ESO 509-IG066 \citep{kosec2017}, MCG+04-48-002/NGC 6921 \citep{koss2016}, and NGC 833/NGC 835 \cite{oda2018}, all of which are local, relatively bright, and have separations large enough for \textit{NuSTAR} to spatially resolve ($\sim 16''-91''$, though at 16'' the two sources would appear as a convolved, elongated single source). However, hard X-ray observations of dual AGNs with \textit{NuSTAR} are relatively uncommon, and most dual AGN studies performed with \textit{NuSTAR} suffer from non-detections due to \textit{NuSTAR}'s sensitivity limit \citep[e.g.,][]{ptak2015,ricci2017MNRAS,pfeifle2023c} and/or convolved sources due to \textit{NuSTAR}'s angular resolution \citep[HPD$\sim58''$, 18'' FWHM;][]{yamada2018,iwasawa2018,pfeifle2019b,iwasawa2020,pfeifle2023c}. This requires a great deal of care and caution when performing the spectroscopic analysis and interpreting the results \citep[e.g., the case of NGC 6240,][]{nardini2017}. To usher in revolutionary advances in hard X-ray dual AGN science, new facilities with higher spatial resolution and greater sensitivity at hard X-ray energies ($>10$\,keV) are required.

In this paper, we discuss the feasibility of hard X-ray dual AGN science with the High Energy X-ray Probe (HEX-P) mission concept. The current mission design for \textit{HEX-P} is described in Section~\ref{sec:missiondesign}. In Section~\ref{sec:simulations} we describe our \textit{HEX-P} imaging and spectroscopic simulations, generated using the \textsc{soxs} \citep{zuhone2023} and \textsc{sixte} \citep{dauser2019} software packages. We outline the potential for resolving hard and soft X-ray signatures from dual AGNs with \textit{HEX-P} using the instrument and facility design described in Section~\ref{sec:HEXPimaging}, the feasibility of using nested sampling and Bayesian statistical source detection techniques to probe closely-separated hard X-ray sources in Section~\ref{sec:soudetect}, and we compare these results to what is currently possible with \textit{NuSTAR} in Section~\ref{sec:compnustar}. In Section~\ref{sec:arp299}, we examine simulated hard X-ray imaging and spectra for the dual AGN in Arp 299 as a test case for \textit{HEX-P} dual AGN science, wherein we use state-of-the-art torus models and Bayesian fitting techniques to constrain the line-of-sight column densities, photon indices, and other spectroscopic parameters of the AGNs. We discuss the issue of spectral contamination for closely separated sources in Section~\ref{sec:spectroscopy}. In Section~\ref{sec:synergies} we briefly outline synergies with upcoming ground-based and space-based facilities. We provide concluding remarks in Section~\ref{sec:conclusions}

\section{Mission Design}
\label{sec:missiondesign}
The \textit{HEX-P} (Madsen et al. 2023, in preparation) is a probe-class mission concept that offers sensitive broad-band spectral coverage (0.2--80\,keV) with exceptional spectral, timing and angular capabilities. It features two high-energy telescopes (HETs) that focus hard X-rays, and one low-energy telescope (LET) that focuses lower energy X-rays.

The LET consists of a segmented mirror assembly coated with Ir on monocrystalline silicon that achieves a half power diameter of 3.5'', and a low-energy DEPFET detector, of the same type as the Wide Field Imager \citep[WFI;][]{meidinger2020} onboard Athena \citep{Nandra2013}. It has 512 x 512 pixels that cover a field of view of 11.3' x 11.3'. It has an effective passband of 0.2--25\,keV, and a full frame readout time of 2\,ms, which can be operated in a 128 and 64 channel window mode for higher count-rates to mitigate pile-up and faster readout. Pile-up effects remain below an acceptable limit of $\sim 1\%$ for a flux up to $\sim 100$\,mCrab (2--10 keV) in the smallest window configuration. Excising the core of the PSF, a common practice in X-ray astronomy, will allow for observations of brighter sources, with a  typical loss of up to $\sim 60\%$ of the total photon counts.

The HET consists of two co-aligned telescopes and detector modules. The optics are made of Ni-electroformed full shell mirror substrates, leveraging the heritage of XMM-Newton \citep{Jansen2001}, and coated with Pt/C and W/Si multilayers for an effective passband of 2--80\,keV. The high-energy detectors are of the same type as flown on \textit{NuSTAR} \citep{harrison2013}, and they consist of 16 CZT sensors per focal plane, tiled 4 x 4, for a total of 128 x 128 pixel spanning a field of view slightly larger than for the LET, 13.4’x13.4’.

The broad X-ray passband and superior sensitivity will provide a unique opportunity to study dual AGNs and galaxy mergers across a wide range of energies, luminosity ratios, pair separations, and dynamical regimes.

All the simulations presented here were produced with a set of response files that represent the observatory performance based on current best estimates as of Spring 2023 (see Madsen+23). The effective area is derived from ray-tracing calculations for the mirror design including obscuration by all known structures. The detector responses are based on simulations performed by the respective hardware groups, with an optical blocking filter for the LET and a Be window and thermal insulation for the HET. The LET background was derived from a GEANT4 simulation \citep{Eraerds2021} of the WFI instrument, and the HET backgrounds was derived from a GEANT4 simulation of the \textit{NuSTAR} instrument. Both simulations adopt the L1 orbit for HEX-P. 

\section{Simulating Dual AGNs Using SOXS and SIXTE}
\label{sec:simulations}
To simulate \textit{HEX-P} observations of dual AGNs, we rely upon two software suites: (1) the \textsc{Simulated Observations of X-ray Sources} \citep[\soxs{},][]{zuhone2023} suite, to create the SIMulated inPUT (SIMPUT) files that store the spatial and spectral models used for the simulations, and (2) the \textsc{Simulation of X-ray Telescopes} \citep[\sixte{},][]{dauser2019} suite, which was used to generate the actual simulated observations. 

Each simulation includes two AGNs represented as point source spatial models (hereafter AGN 1 and AGN 2), and we produced simulations for a variety of separations, ranging from 50'' down to 20'' in increments of 10'' as well as from 20'' down to 0'' in increments of 2''. Here, the 0'' case represents the ``null'' expectation of what a single source would look like at the combined flux level of the AGNs in our simulations and it is included for visual comparison purposes. For illustrative purposes in this work, we also include simulations of AGNs separated by 15'', 5'', 3'', and 1''. We use the following components to develop the spectral models for the AGNs: (1) an absorbed power law ($\Gamma=1.8$) that accounts for photoelectric absorption and Compton scattering; (2) reprocessed emission from a torus \citep[\textsc{borus}, ][which self consistently accounts for absorption, fluorescence, and Compton scattering for a toroidal distribution of obscuring gas]{balokovic2018}; (3) two soft X-ray thermal components (\textsc{apec}) to model emission due to star formation, motivated by the common presence of soft thermal components in many dual AGNs and candidate systems \citep[see, e.g., Section~5.4 in][]{pfeifle2023c}; (4) a power law to model Thomson scattering (0.5\%). This AGN spectroscopic model is described in \textsc{xspec} as:
\vspace{-2mm}
\begin{equation}
tbabs\times(borus+ztbabs\times cabs\times cutoffpl+const\times cutoffpl+apec+apec)
\end{equation}
Each AGN spectrum is simulated at a redshift of $z=0.05$ (D$_{\rm{L}}=222.3$\,Mpc). We assume an approximately edge-on torus viewing angle of $\theta _{\rm{view}}$ = 70$^{\circ}$ and a covering factor of $\rm{C}=85\%$; this latter assumption is motivated by the expectation that mergers should result in heavily obscured AGNs with high covering factors \citep[][]{ricci2017MNRAS}. AGNs 1 and 2 have observed (uncorrected for intrinsic absorption) $2-10$\,keV fluxes normalized to $1.0\times10^{-13}$ and $5.0\times10^{-14}$ erg\,cm$^{-2}$\,s$^{-1}$, respectively. For each choice of separation, we run the spectral model of each AGN over a small list of line-of-sight column densities, log($N_{\rm{H}}/\rm{cm}^2) = 22, 23, 24$ (for simplicity we assume the average torus column density and the line-of-sight column density are equal); we therefore end up with nine SIMPUT files for each separation, where each SIMPUT file probes a separate pairing in terms of obscuration; i.e., one Compton-thick ($N_{\rm{H}}>10^{24}$\,cm$^{-2}$) pair, four Compton-thick and Compton-thin ($N_{\rm{H}}<10^{24}$\,cm$^{-2}$) pairs, and four Compton-thin-Compton-thin pairs. \soxs{} is then used to convolve the spatial and spectral models to produce the SIMPUT files for each AGN pairing.

Each SIMPUT file generated above was fed to \sixte{} in order to generate the \textit{HEX-P} HET and LET event files. For each AGN pairing, we assumed a 50\,ks exposure (which could be the typical exposure time in the \textit{HEX-P} wide area survey, and is also similar in length to archival \textit{NuSTAR} observations of dual AGNs, i.e., $\sim20-60$\,ks) and generated event files for HET 1, HET 2, an effective (co-added) 2-camera HET event file, and a LET event file using the v7 \textit{HEX-P} response files (see Madsen et al. 2023). Science images were generated for all three \textit{HEX-P} cameras using the \textsc{imgev} command and spectra were extracted using the \textsc{makespec} command from 10'' apertures (unless states otherwise in the text) centered on the AGN positions. We performed the same set of commands in \sixte{} for the analogous \nustar{} simulated imaging using the \nustar{} response files \footnote{https://www.nustar.caltech.edu/page/response\_files} and a background file derived from the COSMOS field \citep{civano2015}.

\section{Hard X-ray Imaging Results for Dual AGNs}
\label{sec:hardxrayimaging}

\subsection{HEX-P HET and LET Imaging of Dual AGNs}
\label{sec:HEXPimaging}
\begin{figure}[]
\begin{center}
\includegraphics[width=\linewidth]{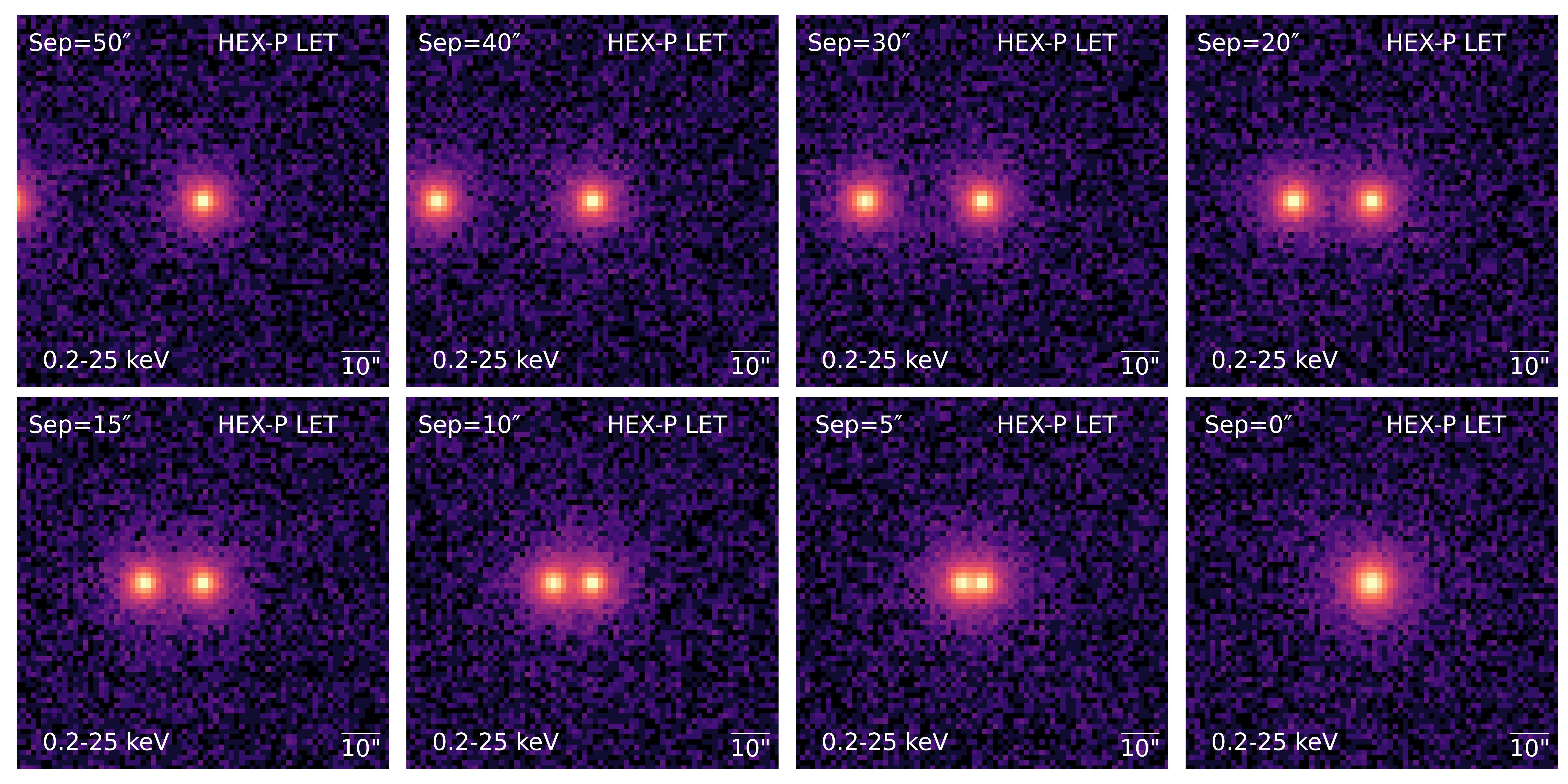}
\end{center}
\caption{This 8-panel figure shows the LET 0.2-25 keV 50 ks images (1-pixel smoothing).  Panels represent angular separations ranging from 50’’ down to 0’’. The two AGNs are distinguishable by eye down to 5'' and can be identified using Bayesian detection methods (see Section~\ref{sec:soudetect}) down to separations of $\sim 2$''. The ``null'' case of 0'' (representing a single AGN at the combined flux level of our sources) is included for visual comparison. The AGNs have observed 2-10 fluxes of $1.0\times10^{-13}$ and $5.0\times10^{-14}$ erg\,cm$^{-2}$\,s$^{-1}$. These panels display the case in which both AGNs are Compton-thick, though the Compton-thick/Compton-thin and Compton-thin/Compton-thin cases show similar X-ray morphologies.}
\label{fig:LET}
\end{figure}

\begin{figure}[]
\begin{center}
\includegraphics[width=\linewidth]{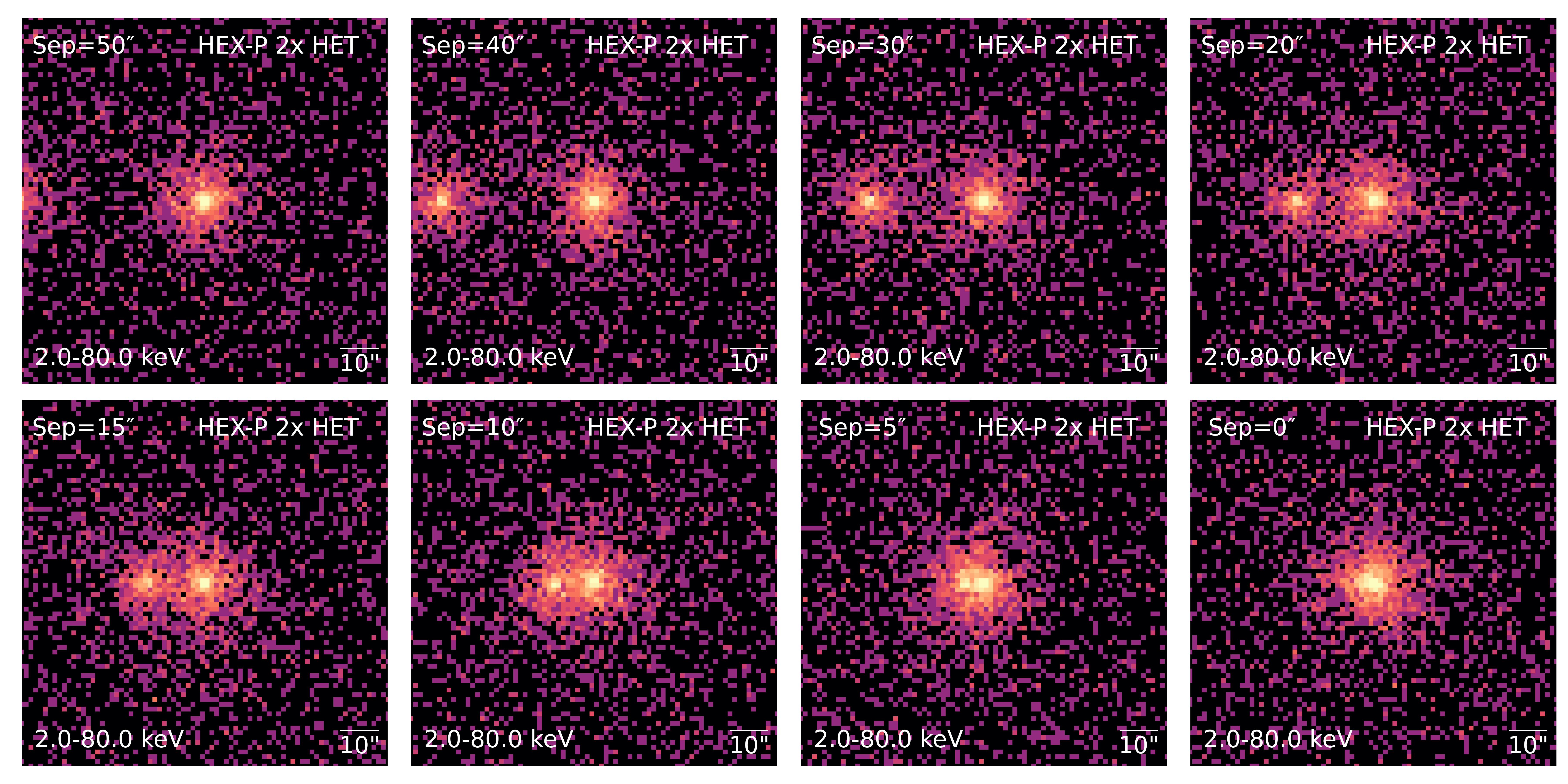}
\end{center}
\caption{This 8-panel figure shows the effective 2-camera HET 2-80 keV, 50 ks images (1-pixel smoothing). As in Figure~\ref{fig:LET}, panels represent angular separations ranging from 50’’ down to 0’’. The two AGNs are distinguishable by eye down to 5'' and can be identified using Bayesian detection methods (see Section~\ref{sec:soudetect} down to separations of $\sim 2$''. The AGNs have observed 2-10 fluxes of $1.0\times10^{-13}$ and $5.0\times10^{-14}$ erg\,cm$^{-2}$\,s$^{-1}$. These panels display the case in which both AGNs are Compton-thick, though the Compton-thick/Compton-thin and Compton-thin/Compton-thin cases show similar X-ray morphologies.}
\label{fig:HET}
\end{figure}

We show a subset of our dual AGN simulation suite in Figure~\ref{fig:LET} and Figure~\ref{fig:HET} for the 0.2-25\,keV LET and (co-added two-camera) 2-80\,keV HET imaging, respectively, with each panel representing a distinct pair separation, ranging from 50'' down to 0'' in 10'' or 5'' intervals. AGN 1 (centered in the field of view) is clearly seen in the center of each panel, while AGN 2 can be seen moving closer in from the left with each successive panel. These images clearly show the two AGNs are distinguishable by eye down to $\sim$5'' in both the HET and LET; we include a finer grid of separations in the Appendix with separations ranging from 5'' down to 1'' in 1'' intervals, which demonstrates that the AGNs can be distinguished by eye even down to $\sim4$'' in HET imaging and $\sim3$'' in LET imaging.

Probing dual AGNs with \textit{HEX-P} down to separations $\lesssim5$'' is a particularly important task, as dual AGNs are predicted to experience the peak of their merger-driven luminosities and obscuration in late-stage mergers \citep{capelo2015,blecha2018} when the nuclei of the galaxies are separated by $\lesssim10$\,kpc ($\lesssim 10.2$'' at $z=0.05$). \textit{HEX-P}'s access to X-ray energies beyond 10\,keV will be vital to constraining the intrinsic luminosities and distinct column densities along the line-of-sight to each AGN in a merging system, an impossibility with \textit{NuSTAR} for all but the nearest and brightest \citep{nardini2017} and/or most widely-separated dual AGNs known \citep[e.g.,][]{ptak2015,kosec2017,oda2018}. Furthermore, broadband coverage is fundamental not only to detect obscured AGN, but also to characterize their obscuration without biases; broadband hard X-ray coverage with \textit{NuSTAR}, for example, has led to dramatic refinements in AGN column density estimates compared to constraints derived from only soft X-ray or joint soft and ultra-hard X-ray coverage \citep[e.g.,][]{marchesi2018,marchesi2019}. Here we discuss the possibility of probing dual AGNs with separations $1''<r_p<5''$ in the hard X-rays with \textit{HEX-P}.

\label{sec:soudetect}
We analyse each simulated event file with a single and double point source model, with wide, uninformative priors on the location and the total counts of each point source, I. The first point source is assumed to be located near the image centre $\pm$10px. Both models also include a background count rate, B, that is constant across the image and left free during the fit. A standard Poisson likelihood is used, with the symmetric PSF model $\mathrm{PSF}(\Delta r)$ extracted from simulations:

\begin{equation}
\mathcal{L}\propto\frac{\prod_{i\in\mathrm{events}}M(x_{i},y_{i}|x_{c},y_{c},I,B)}{\int\int
M(x_{i},y_{i}|x_{c},y_{c},I,B)\,\mathrm{d}x\,\mathrm{d}y}  
\end{equation}

For the single point source model,
\begin{equation}
M(x,y|x_{c},y_{c},I,B)=I\times\mathrm{PSF}\left(\sqrt{\left(x_{i}-x_{c}\right)^{2}+\left(y_{i}-y_{c}\right)^{2}}\right)+B
\end{equation}
while for the two point source model there are two centres and intensities.

The Akaike Information Criterion (AIC; \citealt{akaike1974}) of the two models is compared to determine whether the single or double source model is preferred. In order for this comparison to take place, we first calibrated our detection algorithm using simulations of convolved X-ray sources, where we simulated two co-spatial X-ray sources with identical fluxes to those in our AGN simulations described in Section~\ref{sec:simulations}. We simulated 100 cases of convolved sources for the LET as well as for the HET$\times$2, ran the detection algorithm on each case, and recorded the resulting AIC values. This calibration set the baseline for an acceptable AIC threshold to differentiate between a single and dual source case. We are able to detect X-ray point sources in both the LET 0.2-25\,keV bandpass and the HET 2-80\,keV bandpass down to separations of 2''. This result comes with the caveat that this detection routine is a proof-of-concept and the results are subject to change based on the exposure time and the fluxes of the sources: all else being equal, we cannot necessarily differentiate between multiple sources at 2'' or 3'' separations for lower exposure times; alternatively, we can probe down to closer pair separations ($<2''$) for higher flux levels (a more rigorous examination is left to future works). Nonetheless, this is essentially an order-of-magnitude improvement over current capabilities with \textit{NuSTAR}.

While dual AGNs are predicted to be heavily obscured in late-stage mergers \citep{blecha2018}, there are cases of heavily obscured dual AGNs in widely-separated merging pairs of galaxies \citep{koss2012,derosa2018,derosa2023}. Given that dual AGNs are also predicted to become heavily obscured after only the second pericenter passage \citep{blecha2018}, and that AGNs are expected to flicker on and off across the merger sequence \citep[e.g., the AGN duty cycle $<<$ the merger timescale][]{schawinski2015,goulding2019}, \textit{HEX-P} may reveal multi-modal distributions of heavily obscured AGNs across the merger sequence, rather than only during the latest stages of the merger sequence.

\subsection{A Substantial Improvement over \textit{NuSTAR}}
\label{sec:compnustar}
In an analogous fashion to the \textit{HEX-P} simulated event files displayed in Figures~1 and 2, we have used identical input models, exposure times, and separations to develop \textit{NuSTAR} dual AGN imaging that incorporates the \textit{NuSTAR} background derived from the COSMOS field \citep{civano2015}. Figure~3 displays these simulated \textit{NuSTAR} images (where the color scale matches that of the HET imaging in Figure~\ref{fig:HET}), in which the positions of the AGNs are denoted with black crosses; AGN 1 can be seen in a few (but not all) of the panels above the background while AGN 2 is virtually imperceptible in almost every panel. This figure clearly illustrates that - at these realistic observed flux levels of $1\times10^{-13}$ and $5\times10^{-14}$ erg cm$^{-2}$ s$^{-1}$ - \textit{NuSTAR} cannot reliably identify dual AGNs in these pairings; it is important to note that dual AGNs separated by $\sim20$'' and $0$'' are essentially indistinguishable, emphasizing that \textit{NuSTAR} cannot be used to reliably resolve $<20$’’ dual AGNs. \textit{HEX-P}, on the other hand, can spatially resolve each AGN in both the HET and LET imaging down to separations of $\sim5$'' (and $\sim2''$ when rigorous detection algorithms are used; see Section~\ref{sec:soudetect}); this is a substantial improvement over \textit{NuSTAR} and emphasizes that the future of hard X-ray dual AGN science relies critically upon access to facilities such as \textit{HEX-P}.

\begin{figure}[]
\begin{center}
\includegraphics[width=\linewidth]{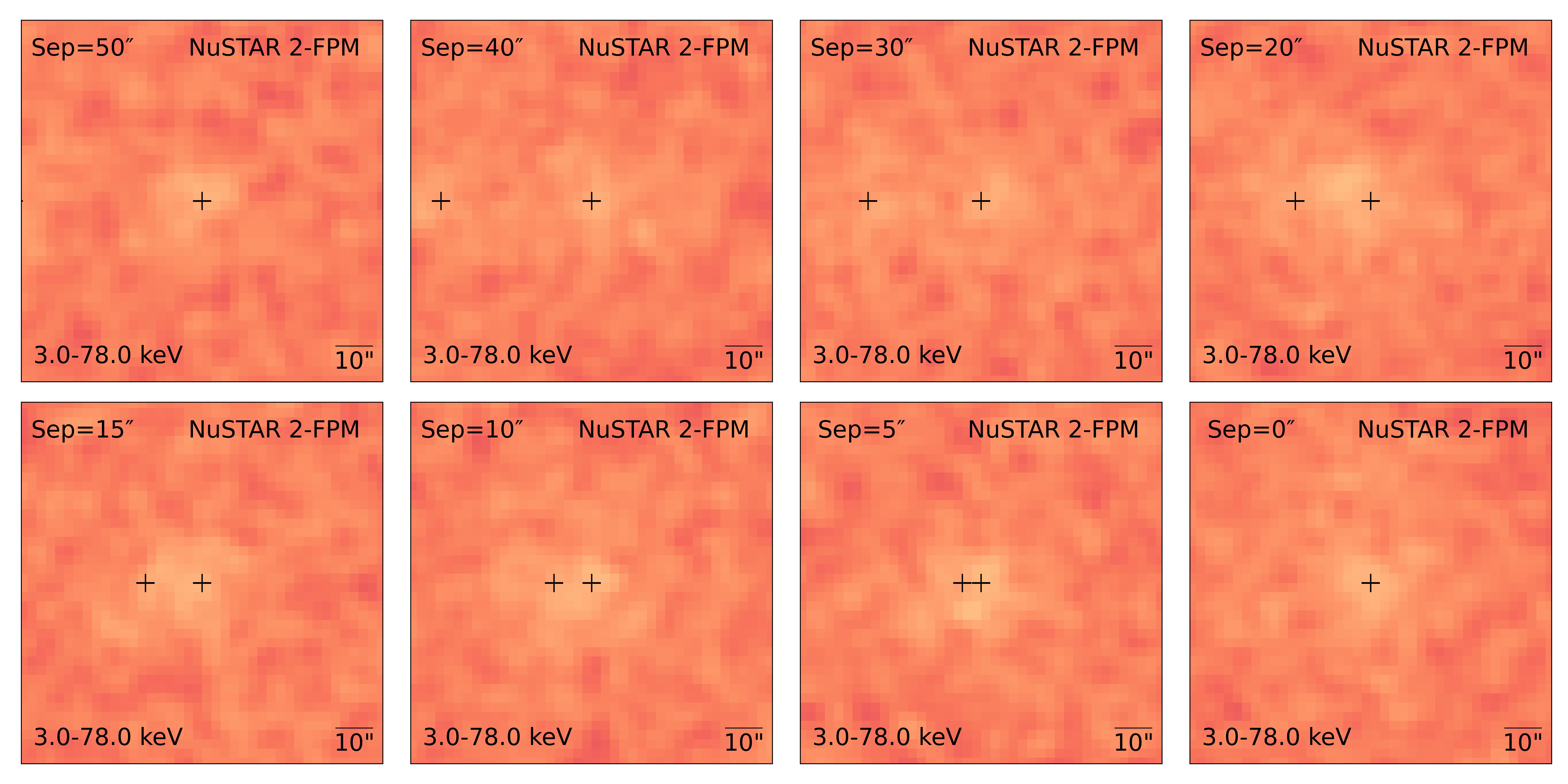}
\end{center}
\caption{This 8-panel figure shows simulated two-camera 3-78 keV FPMA$+$B 50\,ks imaging with \textit{NuSTAR} for the same suite of dual AGNs described in Section~\ref{sec:simulations}. The suite of AGNs range in separation from 50'' down to 0''; the scale bar in the bottom right corner of each panel indicates an angular scale of 10''. These panels use the same scaling as in Figure~\ref{fig:LET} for the \textit{HEX-P} HET 2-80 keV imaging. The primary AGN can be seen above the background, but the weaker secondary AGN is essentially imperceptible above the background in all of the panels. Black markers are included to denote the positions of the two AGNs.}
\label{fig:3}
\end{figure}

\subsection{A Substantial Improvement over Soft X-ray Missions}
\begin{figure}[]
\begin{center}
\includegraphics[width=\linewidth]{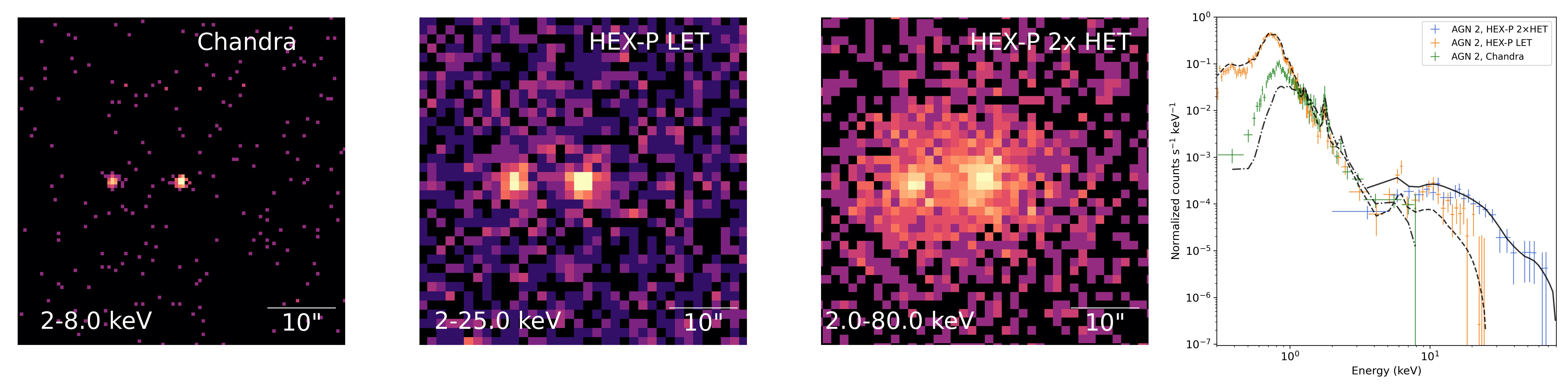}% This is a *.eps file
\end{center}
\caption{This four-panel figure shows a comparison between simulated Chandra, \textit{HEX-P} LET, and \textit{HEX-P} HET imaging and spectroscopy. The three X-ray imaging panels show the \textit{Chandra} 2-8\,keV, \textit{HEX-P} LET 2-25\,keV, and \textit{HEX-P} 2-80\,keV energy bands. The fourth panel displays the extracted \textit{Chandra} (green) and \textit{HEX-P} LET (orange) and HET (blue) spectra for AGN 2 (the weaker of the two AGNs, see Section~\ref{sec:simulations}), binned at 5$\sigma$. }
\label{fig:chandracomp}
\end{figure}

In Figure~\ref{fig:chandracomp} we display a simulated 50\,ks \textit{Chandra} 2-8\,keV image alongside the 2-25\,keV LET and 2-80\,keV HET$\times$2 imaging for the case where the AGNs are separated by 10''. While \textit{Chandra} can clearly resolve two X-ray cores emitted by the AGNs at $>2$\,keV, the detections are obviously limited to energies $<8$\,keV. To offer a quantitative comparison between \textit{Chandra} and \textit{HEX-P}, we extracted (background subtracted) counts for each AGN in the \textit{Chandra}, LET, and HET imaging. The AGNs are detected with $175\pm14$ counts (AGN 1) and $67\pm9$ counts (AGN 2) in the 2-8\,keV \textit{Chandra} band when using 3'' extraction apertures ($\sim100\%$ EEF); it is important to note that most dual AGNs and candidates are observed for $<30$\,ks, so with a more realistic exposure time these sources would only have been detected with $\lesssim100$ and $\lesssim40$ counts. Contrasting this with the detections in LET, we detect $215\pm16$ and $129\pm13$ counts in the LET 2-25\,keV band for the two AGNs when using 3.5'' extraction apertures ($\sim50\%$ EEF). At this separation, the two AGNs begin to severely contaminate one another in the HET imaging when using 8.5'' (50\% EEF) apertures, so we instead excise only the inner 3.5'' of each hard X-ray source (20-25\% EEF): we find the two AGNs are detected in the HET with $278\pm17$ and $189\pm14$ counts in the 2-80\,keV band; these comparisons demonstrate that, given the same exposure time, \textit{HEX-P} offers comparable, if not superior, photometric analyses on top of access to the hard X-rays. 

To compare the spectroscopic capabilities of \textit{HEX-P} to \textit{Chandra}, we extracted the spectrum of the weaker of the two AGNs (AGN 2) from the \textit{Chandra} and \textit{HEX-P} images using a 3'' aperture for the \textit{Chandra} image, a 3.5'' aperture for LET, and a 3.5'' aperture in HET$\times$2; these apertures represent the $\sim100$\%, $\sim50$\%, and $22$\% EEF for each instrument. This comparison therefore represents the ``best-case'' scenario for a \textit{Chandra} exposure, a ``standard case'' for a LET exposure, and a ``poor case'' scenario for HET$\times$2. The spectra (Chandra: green points, LET: orange points, HET$\times$2: blue points) are shown in the fourth panel of Figure~\ref{fig:chandracomp} and binned at 3$\sigma$ per bin for visualization purposes. Only LET and HET can access energies beyond 10\,keV, where HET can reach out even to 80\,keV. The Compton-hump beyond 10\,keV is a spectral feature critical for accurate constraints on the line-of-sight absorbing columns (e.g., P. G. Boorman, in preparation), and it is clearly accessible to the HET. The Fe K$\alpha$ emission line is another crucial spectral imprint of cold reflection, and a simple fit using a phenomenological power law model reveals that the addition of a Gaussian emission line to account for Fe K$\alpha$ does not statistically improve the fit to the simulated \textit{Chandra} spectrum, but it \textit{does} statistically improve the fit to the LET spectrum. Thus, this case study illustrates clearly that \textit{HEX-P} can simultaneously access two AGN reflection features important for accurate column density constraints that would, in this case, be inaccessible to soft X-ray missions. It is important to note, too, that while soft X-ray missions can in principle detect the Fe K$\alpha$ emission line, intrinsic parameter estimations (including column density determination) can still be degenerate when using the Fe K$\alpha$ emission line alone \citep[e.g.,][]{lamassa2017} without harder X-ray features, whereas constraints that use the Fe K$\alpha$ line and the Compton-hump in tandem offer clearer and more reliable constraints \citep[e.g.,][]{lamassa2019}. Without simultaneous access to these X-ray spectral features, soft X-ray missions will continue to be at a disadvantage when it comes to answering open questions related to dual AGN obscuration and fueling in the X-rays.

\subsection{Hard X-ray Dual AGN Science as a Function of Redshift}

\begin{figure}[]
\begin{center}
\includegraphics[width=1.0\linewidth]{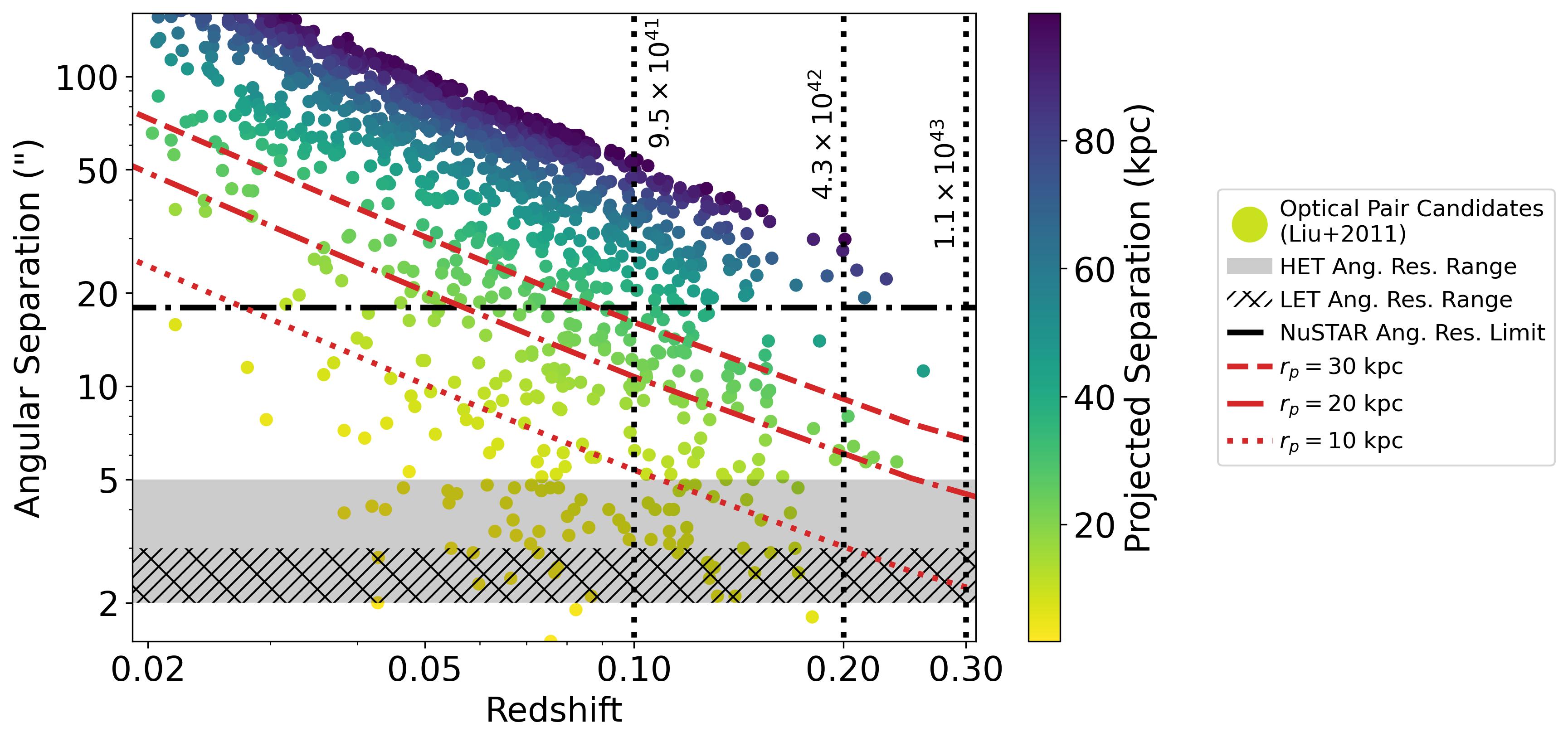}
\end{center}
\caption{Redshift vs. angular separation.  We have plotted the sample of optical dual AGN candidates from \citet{liu2011} at $z<0.3$. Markers are color coded according to the projected separation of the pair (given on the auxiliary axis). Red dashed, dash dotted, and dotted lines indicate projected separations of 30, 20, and 10 kpc. The grey, shaded region indicates the HET angular resolution limits when examining sources visually by eye ($\sim5''$) and when using Bayesian detection algorithms ($\sim2''$, see Section~\ref{sec:soudetect}); the hashed and shaded region similarly indicates the LET angular resolution limits ($\sim3''$ by eye, $\sim2''$ using Bayesian detection routines). The angular resolution limits derived via the Bayesian detection algorithm in Section~\ref{sec:soudetect} are subject to change based upon the exposure time and source fluxes; here they refer specifically to the 50\,ks exposure time and source fluxes listed in Section~\ref{sec:simulations}. \textit{NuSTAR}'s angular resolution limit \citep[18'' PSF,][]{harrison2013} is denoted with a dashed-dotted black line. Vertical, dotted lines denote the observed 2-10\,keV AGN luminosity that corresponds to the \textit{HEX-P} flux sensitivity limit of $3.8\times10^{-14}$\,erg\,cm$^{-2}$\,s$^{-1}$ in a 50\,ks exposure (signal-to-noise of 5).}
\label{fig:angvredshift}
\end{figure}

Current hard X-ray facilities (e.g., \textit{NuSTAR}) limit the study of distinct AGNs within dual AGN pairings to only the nearest, brightest, and/or more widely-separated systems. Figure~\ref{fig:angvredshift} shows the distribution of angular separations for optical dual AGN candidates from \citet{liu2011} as a function of redshift out to $z\sim0.3$. The overlaid black dash-dotted line indicates the angular resolution of \textit{NuSTAR} \citep[18'' FWHM, 58'' HPD,][]{harrison2013}, while the red dashed, dash-dotted, and dotted curves indicate nuclear pair separations of 30, 20, and 10 kpc. Overlaid on this figure, we have also plotted the corresponding \textit{HEX-P} 2-10\,keV luminosity limits (vertical dotted, black lines) as a function of redshift, informed by the limiting flux of $3.8\times10^{-14}$\,erg\,cm$^{-2}$\,s$^{-1}$ for a 50\,ks exposure; the vast majority of dual AGNs have 2-10\,keV luminosities $\lesssim10^{43}$\,erg\,s$^{-1}$. Therefore, $z\approx0.3$ is an approximate redshift limit for dual AGN studies with \textit{HEX-P}. Even late-stage mergers ($\lesssim10$\,kpc) become virtually inaccessible to \textit{NuSTAR} already by $z\approx0.025$'', and more widely separated pairs ($\sim30$\,kpc) become inaccessible by only $z\approx0.09$. \textit{HEX-P}, on the other hand, thanks to resolving separations of $\sim 5$'' (possibly even $\sim2''$), can distinguish AGNs in late-stage mergers out to redshifts of $z\sim0.1$ (or even $z\sim0.2$ with Bayesian analyses, Section~\ref{sec:soudetect}) and can access distinct AGNs in $\sim30$\,kpc-separated duals out to $z\sim0.3$. Though \textit{distinct} AGNs within late-stage mergers will not be probed beyond $z\sim0.1-0.2$, \textit{HEX-P} will be able to probe the `gross' X-ray properties of these systems\footnote{Selected presumably in the soft X-rays with \textit{Chandra}, via high-resolution radio imaging, space-based high-resolution optical or near-IR spectroscopy (e.g., Hubble Space Telescope or JWST), or via ground-based adaptive-optics-assisted optical or near-IR observations. However, some fraction of dual AGNs will be missed at optical wavelengths \citep[e.g.,][]{koss2012} and require the detection of distinct X-ray AGNs for unambiguous confirmation.}; even when unresolved, \textit{HEX-P} is required to probe the Compton hump and understand the circumstances of the nuclear environment properly, which cannot be done using soft X-ray, optical, near-IR, or radio observations.  \textit{HEX-P} will probe distinct AGNs in early-stage mergers well beyond $z\sim0.1$ and aid in our understanding of dual AGN activation and obscuration across the merger sequence.

\section{A Test Case for HEX-P: Arp 299}
\label{sec:arp299}
One illustrative case study for examining the effectiveness of \textit{HEX-P} over \textit{NuSTAR}, in terms of spectroscopic and imaging constraints, is Arp 299, a local (44 Mpc, or $z\sim0.01$) dual AGN comprising an X-ray bright and heavily obscured AGN in the southwest nucleus (Arp 299-B, detected by \textit{Chandra} and \textit{NuSTAR}, \citealp{ballo2004,ptak2015}) and an AGN in the northeast nucleus (Arp 299-A) identified as a flat spectrum radio source \citep{pereztorres2010}, and tentatively identified via mid-IR SEDs \citep{alonsoherrero2013} and ionized Fe K$\alpha$ emission \citep{ballo2004}. As reported by \citet{ptak2015}, Arp 299-B is detected by \textit{NuSTAR} with a hard X-ray flux of $F_{10-30\,keV}=3.5\times10^{-12}$ erg cm$^{-2}$ s$^{-1}$, while Arp 299-A is dominated by X-ray binaries below 10\,keV and contains no hard X-ray emitting AGN above a $10-30$\,keV limit of $<3.5\times10^{-13}$ erg cm$^{-2}$ s$^{-1}$. Here we demonstrate that \textit{HEX-P} would not only clearly detect both of the reported AGNs in Arp 299 in the 0.2-80\,keV band but would spatially resolve the AGNs in both HET and LET imaging. 

Our input models are informed by the model choices and spectral fitting results in \citet{ptak2015}. Specifically, we adopt the following model to describe the spectrum of each nucleus: 
\begin{equation}
    tbabs\times(borus+ztbabs\times cabs\times zcutoffpl+zcutoffpl)
\end{equation}
where we have used the \textsc{borus02} model \citep{balokovic2018} rather than MYTorus, which was used by \citet{ptak2015}. This model incorporates host galaxy absorption (\textsc{tbabs}), reflected emission (\textsc{borus02}), an absorbed intrinsic power law that accounts for photoelectric absorption and Compton scattering, and a cut-off power law to account for soft X-ray emission from X-ray binaries. For Arp\,299-B, we assume $\Gamma=1.95$, $N_{\rm{H}}=4\times10^{24}$\,cm$^{-2}$ (for simplicity we assume the average torus column density and the line-of-sight column density are equal), covering factor cos($\theta_{\rm{open}})=0.85$, and we normalize the $10-30$\,keV emission to $3.5\times10^{-12}$ erg cm$^{-2}$ s$^{-1}$. For Arp\,299-A, we assume $\Gamma=1.8$, $N_{\rm{H}}=5\times10^{22}$\,cm$^{-2}$ \cite[since it is likely an unobscured, low luminosity AGN, e.g.,][]{ptak2015}, and covering factor cos($\theta_{\rm{open}})=0.5$; while \citep{ptak2015} placed an upper limit of $<3.5\times10^{-13}$ erg cm$^{-2}$ s$^{-1}$ on the $10-30$\,keV flux of Arp 299-A ($<10$\% of the observed flux of Arp 299-B), we conservatively assume that Arp 299-A (here, AGN 2) exhibits an observed $10-30$\,keV flux of $7\times10^{-14}$ erg cm$^{-2}$ s$^{-1}$ (i.e. 2\% of the $10-30$\,keV observed flux of Arp 299-B). For both AGNs 1 and 2, we assume $\Gamma=1.8$ and a cutoff energy of 7\,keV for the power law component driven by X-ray binaries; for simplicity, we assume these components contribute equally to the total 2-10\,keV XRB-driven flux ($8.1\times10^{-13}$ erg cm$^{-2}$ s$^{-1}$) found by \citet{ptak2015}, and we therefore normalize both components to $4.05\times10^{-13}$ erg cm$^{-2}$ s$^{-1}$ in the 2-10\,keV band. For our spatial models, we adopt the positions given in Table~1 of \citet{ptak2015}. We once again simulated these AGNs using \textsc{soxs} for SIMPUT file creation and \textsc{sixte} for event file and spectroscopic data product creation.

\begin{figure}
    \centering
    \includegraphics[width=1.0\linewidth]{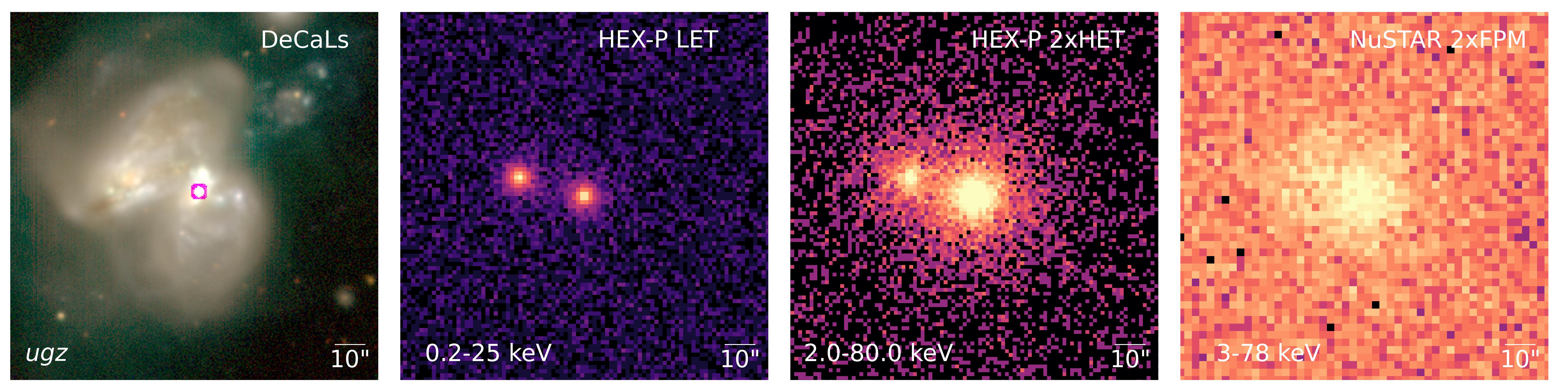}
    \caption{The Arp 299 case study: optical $ugz$ imaging from DeCaLs (left panel), simulated 0.2-25 keV LET and 2-80\,keV HET (2-camera) \textit{HEX-P} imaging (middle panels), and observed 3-78\,keV \textit{NuSTAR} imaging of Arp 299. The energy band or filter is indicated in the bottom left corner of each panel, while the facility or image source resides in the top right corner of each panel. Scale bars in the bottom right corners indicate 10''. The LET and HET images were simulated following the procedure in Section~\ref{sec:arp299} and include spectral and spatial models for the X-ray binary populations and the two AGNs in the merger system. For simplicity, we assume the X-ray binary populations are limited to the nuclei and model them as point sources coincident with the nuclei rather than as extended sources. While the two AGNs cannot be clearly detected in the \textit{NuSTAR} imaging (nor was Arp 299-A detected above 10 keV), the two AGNs can be clearly spatially resolved by both the LET and HET.}
    \label{fig:arp299}
\end{figure}

We show the simulated \textit{HEX-P} LET and HET images in Figure~\ref{fig:arp299} juxtaposed with the tricolor optical $ugz$ imaging from the Dark Energy Camera Legacy Survey (DeCaLs) Legacy Viewer\footnote{https://www.legacysurvey.org/viewer} and simulated \textit{NuSTAR}/FPMA imaging. At $21$'' separation, the soft ($<10\,$keV) X-ray components in the \textit{NuSTAR} imaging were spatially blended, and above 10\,keV the secondary nucleus is not detected by \textit{NuSTAR} \citep[see Figure~6 in ][]{ptak2015}. However, \textit{HEX-P} clearly resolves the two nuclei in both the LET and HET imaging and detects the simulated low luminosity AGN in Arp 299-A at energies beyond 10\,keV.

To demonstrate \textit{HEX-P}'s spectroscopic capabilities, we fit LET and HET spectra extracted with 8'' and 18'' radius apertures, respectively. These aperture choices represent an enclosed energy fraction (EEF) of $\sim100\%$ and naturally suffer from cross-contamination; in practice, aperture sizes that represent only 50\%-60\% EEF are more commonly used when extracting X-ray spectra and would suffer from less cross-contamination, but our aperture choices provide an additional test for \textit{HEX-}P's spectroscopic capabilities in discerning the properties of the distinct AGNs when spectral contamination is present. Arp\,299 represents a prime example in which the LET spectra will be uncontaminated while each HET spectrum will contain contributions from both AGNs, which are separated by 21''. Here we show how the LET can be used to improve the reliability of HET spectral fitting.

We perform the fitting of our simulated Arp\,299 \textit{HEX-P} spectra using v2.9 of the Bayesian X-ray Analysis (BXA; \citealt{Buchner14,BXAsoftwarepaper}) software package, which connects PyXspec \citep{Arnaud1996} to the nested sampling \citep{Skilling2004,Buchner2021c} algorithm MLFriends \citep{buchner2014,Buchner2019} implemented in UltraNest \citep{UltraNest2021JOSS}. Our reasoning for using nested sampling here is two-fold. First, the global parameter exploration enables the traversal of high dimensionality parameter spaces in an efficient manner that is minimally affected by local minima. Second, the pre-defined convergence criteria for nested sampling ensures our fitting process is devoid of biases that can be imposed when fitting of simulated data is not performed in a blind manner (see e.g., \citealt{Kammoun2022}).

The resolving ability of the LET ensures negligible cross-contamination in the spectral extraction of both sources. However, despite the angular resolution of the HET being a considerable improvement on \textit{NuSTAR} (see Figure~\ref{fig:HET}), some cross-contamination remains (see also Figure~\ref{fig:contam}). For this reason, we chose to fit the simulated LET and HET spectra for both AGN 1 and AGN 2 simultaneously whilst incorporating nuisance parameters to describe the amount of flux contamination in each source. Our spectral model thus includes the main spectral components used to simulate the AGNs, as well as a multiplicative factor to each AGN component to account for the relative contamination in each AGN HET$\times$2 spectrum. We manually set the cross-contamination to zero for the LET spectra, since any contamination is negligible, and assume that each HET$\times$2 spectrum contains 100\% of the AGN emission that the extraction region was centered on and some fraction of the other AGN flux contaminating that extraction region.

We assigned log-uniform priors for all line-of-sight absorption, cross-contamination factors, high-energy cut-offs and normalizations. We additionally created custom cos-uniform priors for all variable obscurer geometric angles (namely the torus opening angle and line-of-sight inclination for both AGNs). Finally, the AGN photon indices were assigned Gaussian priors with mean 1.9 and 0.15 standard deviation and the X-ray binary photon index was assigned a uniform prior.

\begin{figure}
    \centering
    \includegraphics[width=1.0\linewidth]{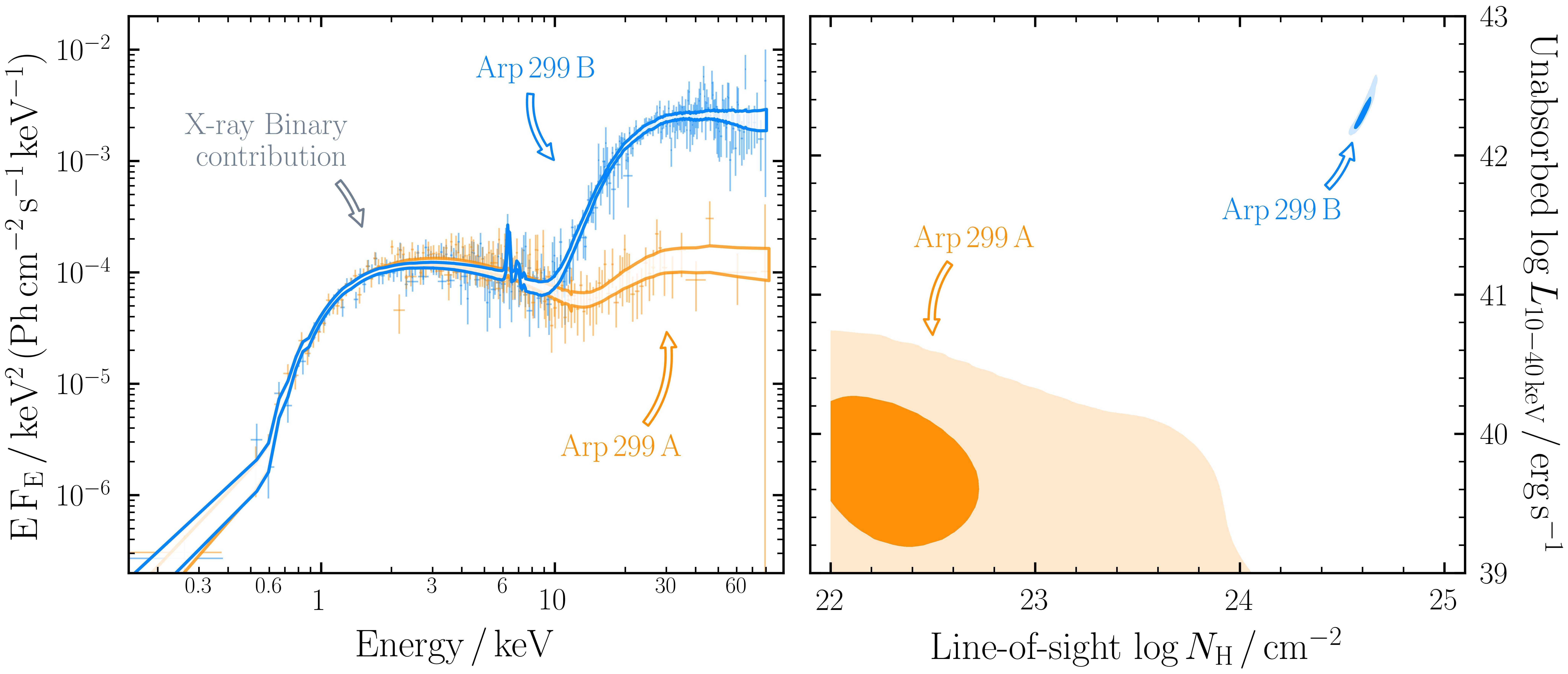}%\vspace{-1cm}
    \caption{Simulated \textit{HEX-P} HET (2-camera, co-added) and LET spectra, and the best-fitting models for the simulated Arp 299-A and Arp 299-B sources. The left panel shows the spectra unfolded with the best-fitting models while the right panel shows the contours for the best-fitting line-of-sight column densities and unabsorbed 10--40\,keV luminosity in units of erg\,s$^{-1}$. Blue and orange data points and model curves correspond to the Arp 299-B and Arp 299-A, respectively. While we find a slightly harder power law slope for the X-ray binary power law component, we recover very similar values for the AGN power law slopes and column densities to those used in the input models.}
    \label{fig:spectroscopicfits}
\end{figure}

Figure~\ref{fig:spectroscopicfits} presents the fit acquired with BXA using a total of 18 free parameters. Owing to the combined broadband, sensitive coverage with the LET and HETs, we constrain the overall intrinsic X-ray luminosities and line-of-sight obscuration (log[$N_{\rm{H}}/\rm{cm}^{-2}$]=$24.60^{+0.04}_{-0.04}$ for Arp-299B and log[$N_{\rm{H}}/\rm{cm}^{-2}$]=$22.43^{+0.64}_{-0.29}$ for Arp-299A) for both AGN well. We additionally note that owing to the simultaneous broadband coverage of the LET and HETs, the spectral fit is additionally devoid of any variability concerns across the full $\sim$0.2--80\,keV passband. By using BXA, we additionally recover the complex posterior degeneracies intricately associated with various system parameters, e.g., torus covering factors/opening angles (log[$\theta_{\rm{open}}$]$=0.88^{+0.08}_{-
0.13}$ for Arp-299B and log[$\theta_{\rm{open}}$]$=0.54^{+0.30}_{-
0.29}$ for Arp-299A), photon index ($\Gamma=1.94^{+0.06}_{-0.10}$ for Arp-299B and $\Gamma=1.89^{+0.14}_{-0.13}$ for Arp-299A), and intrinsic coronal normalisation, the latter two of which are important when propagating uncertainties into intrinsic luminosity estimates or Eddington ratios. These results clearly demonstrate our ability to reliably recover the intrinsic properties for each AGN despite the cross-contamination in the HET imaging.

\section{Spectroscopic Fitting of Dual AGNs}
\label{sec:spectroscopy}
\begin{figure}
    \centering
    \includegraphics[width=0.5\linewidth]{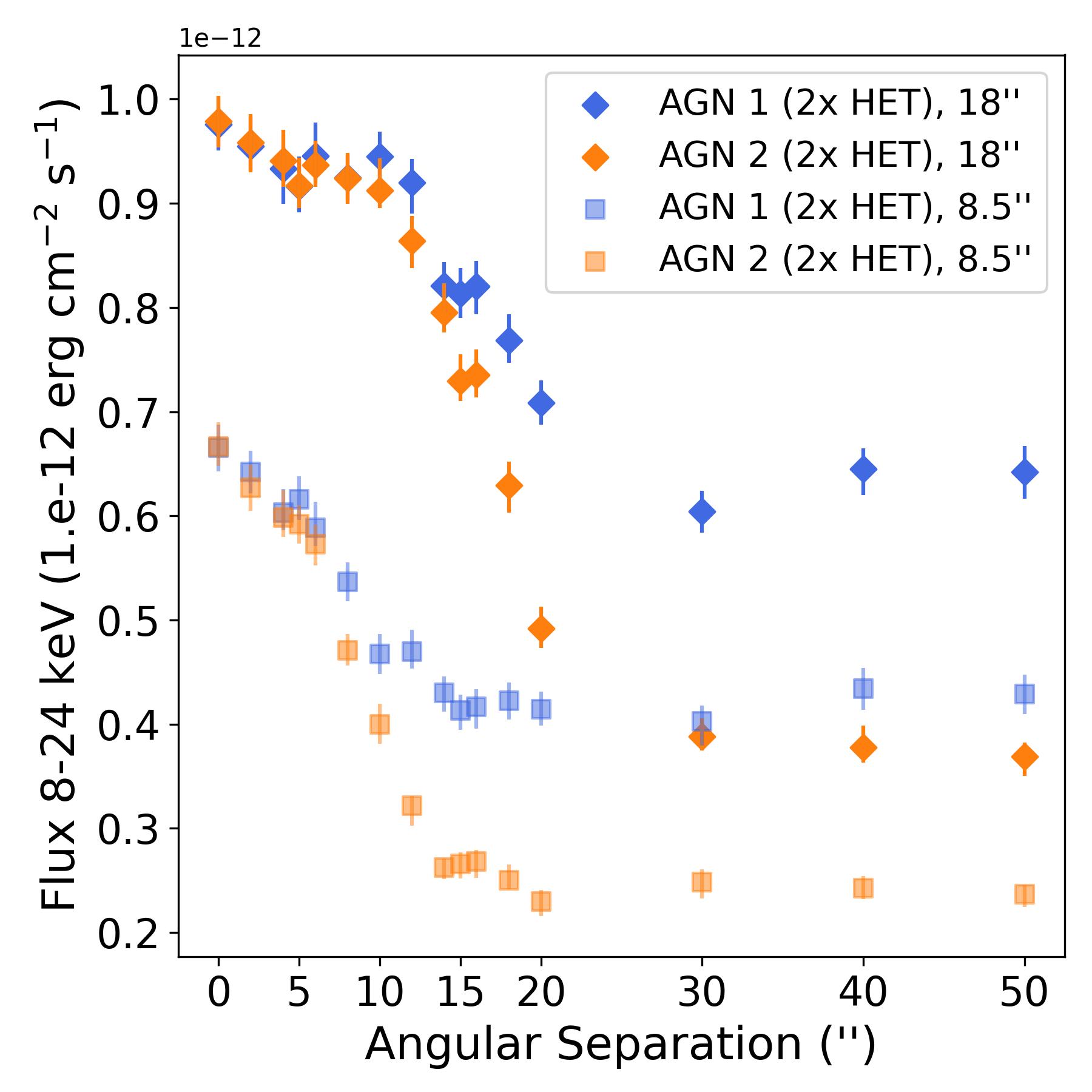}
    \caption{8-24 keV observed fluxes from AGN 1 and 2 (from our general suite of simulations described in Sections~\ref{sec:simulations} and \ref{sec:HEXPimaging}) as a function of pair separation. Diamonds correspond to spectra extracted from 2-camera HET imaging using 18’’ radius apertures ($\sim100\%$ EEF), while squares correspond to spectra extracted from 2-camera HET imaging using 8.5'' radius apertures ($\sim50\%$ EEF); these spectra were fit in \textsc{xspec} using the same input spectral model used to simulate the data, where only the normalizations were free to vary in order to recover the observed flux. While the sharp increase in spectral contamination (traced by the increasing flux of each AGN) at $\sim18$'' for the larger apertures is due to our choice of aperture size, the sharp increase in contamination observed at $\sim10''$ for the smaller apertures is more likely due to the size of the HET PSF itself.}
    \label{fig:contam}
\end{figure}

Simultaneous fitting of soft (\textit{Chandra} and/or \textit{XMM-Newton}) and hard (\textit{NuSTAR}) X-ray spectra has been successfully performed in the past for a few closely separated dual AGN \citep[e.g., NGC 6240, Mrk 273, Mrk 266][]{nardini2017,iwasawa2018,iwasawa2020}. At these pair separations ($\sim 10$'' or lower), careful modeling using tools like BXA \citep[][see above]{buchner2014} should be employed, as the spectra will be convolved at these pair separations and will contain contributions from both AGNs. We illustrate the level of flux contamination in Figure~\ref{fig:contam}, which shows the estimated HET flux for each AGN as a function of pair separation, extracted from the co-added 2xHET images using 18'' radius apertures ($\sim100\%$ EEF; diamonds) and 8.5'' apertures ($\sim50\%$ EEF; squares) for each AGN. As the pair separation decreases, the two AGNs naturally contribute more to each other’s observed flux as a result of the $\sim 17.5$'' HPD of the HET. The large jump in the observed fluxes at angular separations smaller than $\sim18''$ extracted from the 18’’ radius ($\sim100\%$ EEF) apertures, particularly for the weaker AGN, is due to the aperture size rather than the HET PSF. When 8.5'' radius apertures are used ($\sim50\% $ EEF), we find that spectral contamination appears to be modest at separations of $>12''$ and becomes more significant at separations $<12''$, despite our ability to visually resolve the two AGNs in the imaging. The LET, on the other hand, will offer relatively contamination-free spectra down to closer pair separations ($\lesssim6''$). For this reason, it will become increasingly important at smaller pair separations to fit LET and HET AGN spectra simultaneously, because the LET spectra will bolster the accuracy of the X-ray spectral parameters derived from HET spectra.

\section{Synergies with Upcoming Facilities}
\label{sec:synergies}
Given that dual AGNs are predicted to exhibit extremely red infrared colors \citep{blecha2018} in concert with high absorbing columns \citep[e.g.,][]{capelo2015,blecha2018}, infrared observations are a natural choice to pair with hard and soft X-ray imaging and spectroscopy from \textit{HEX-P}. \textit{Euclid}, the recently launched ESA near-IR and optical imaging and spectroscopy mission \citep{laureijs2011}, will provide sub-arcsecond near-IR and optical imaging and near-IR spectroscopy across approximately one third of the sky by the time \textit{HEX-P} launches. By the very nature of its mission design, \textit{Euclid} will act as a new dual AGN survey facility, forming an enormous archival sample of interacting and merging galaxies from which promising near-IR dual AGNs can be selected and followed-up with \textit{HEX-P}. \textit{JWST} will also offer unprecedented near- and mid-IR imaging and spectroscopic observations of small samples of dual AGNs (and candidates) found both within deep surveys \citep[e.g.,][]{comerford2009,civano2010,civano2012} and through targeted observations, and will provide the opportunity to probe not only heavily obscured AGNs but also heavily obscured star formation in, for example, dual AGNs hosted by LIRGs and ULIRGs \citep[e.g.,][]{komossa2003,iwasawa2011,torres-alba2018}. On the other hand, Extremely Large Telescopes (ELTs) -- expected to come online in the 2030s -- will usher in vast improvements in ground-based near-IR and optical spectroscopy and imaging and will experience substantially lower observational overheads and lower subscription rates than \textit{JWST}. ELTs will therefore be paramount for targeting large samples of dual AGNs and may, like \textit{Euclid}, represent the optimal dual AGN survey facilities that can work in tandem with \textit{HEX-P}. In the meantime, high cadence, high resolution ($\lesssim1$'' angular resolution) optical imaging with the upcoming Vera C. Rubin Observatory \citet{ivezic2019} -- in concert with presently available optical and near-IR data from Euclid -- will also aid in the development and preliminary analyses of large samples of dual AGN candidates. The \textit{Nancy Grace Roman Telescope}, slated for launch in the 2030s, will operate at infrared wavelengths with a FOV 100$\times$ larger than \textit{HST} with imaging sensitivity and PSF sizes superior to both \textit{HST} and \textit{Euclid}. \textit{NGR}'s combination of FOV and wavelength coverage will make it an optimal facility for serendipitous dual AGN searches and studies and will complement both \textit{Euclid} and \textit{HEX-P}. Another important upcoming facility will be the `next-generation VLA’ \citep[ngVLA, e.g.,][]{burkespolaor2018}, which will boast 10$\times$ the sensitivity of the Jansky VLA and ALMA with 30$\times$ longer baselines that will enable 0.01''-0.001''  resolution across the 1.2-116\,GHz frequency range. Since radio is also immune to obscuration, the ngVLA will have the sensitivity and resolution to detect dual AGNs to $z\sim1$, if they are sufficiently luminous in the radio. \textit{HEX-P} and the ngVLA could work as separate follow-up facilities, or \textit{HEX-P} could work in tandem with the ngVLA to confirm dual AGNs and measure the obscuring columns and try to constrain torus parameters (the latter of which the ngVLA would not be able to do). A joint HEX-P/ngVLA coverage strategy may be able to simultaneously circumvent X-ray and radio selection biases. In a similar vein, \textit{HEX-P} observations in combination with ALMA 100-200\,GHz continuum observations \citep{ricci2023} could offer unprecedented views to the most heavily obscured dual AGNs. Due to the reduced opacities in these ALMA bands, one can probe through very high column densities.

\section{Conclusions}
\label{sec:conclusions}
The \textit{HEX-P} Probe-class mission concept combines high spatial resolution X-ray imaging ($\sim17.5$'' HPD HET, $\sim3.5$'' HPD LET) and broad spectral coverage (0.2-80 keV) with superior sensitivity to faint fluxes than current facilities and will usher in revolutionary new insights into dual AGN growth across the merger sequence. Hard X-ray imaging and spectroscopy is particularly important to the field of dual AGN science, as these systems are expected to be heavily obscured \citep[e.g.,][]{komossa2003,bianchi2008,mazzarella2012,capelo2015,blecha2018,pfeifle2019a}, making optical and soft X-ray missions biased against their detection and accurate analysis; \textit{HEX-P} is the only Probe mission concept that can fulfil this scientific need. In this work, we have used a suite of dual AGN simulations - developed using \textsc{soxs} \citep{zuhone2023} and \textsc{sixte} \citep{dauser2019} - to demonstrate the following:
\begin{itemize}
    \item Without the aid of specialized detection algorithms, the spatial resolution of \textit{HEX-P} LET and HET will allow us to resolve dual AGNs down to $\sim3''$ and $\sim5''$, respectively ($5''$ corresponds to $\sim4.9$\,kpc at $z=0.05$). NuSTAR, on the other hand, cannot spatially resolve dual AGNs below $\sim18''$ ($\sim17.9$\,kpc at $z=0.05$).
    \item With the aid of Bayesian detection algorithms like that employed in this work, \textit{HEX-P} will be able to confirm dual AGNs down to separations of $\sim2$'' ($\sim2.0$\,kpc at z=0.05) in both HET and LET imaging. The exact spatial resolution limit is determined by a combination of source flux levels and observation exposure time.
    \item Given \textit{HEX-P}'s spatial resolution for both the LET and HET, \textit{HEX-P} will have the capability to probe dual AGNs in late-stage mergers ($<10$\,kpc) out to $z\sim0.1$, and intermediate-stage mergers ($\sim30$\,kpc) out to redshifts of $z\sim0.3$. This is a dramatic improvement over \textit{NuSTAR}, which cannot probe dual AGNs with even 50\,kpc separations beyond $z\sim0.25$.
    \item \textit{HEX-P} will enable distinct (relatively contamination free) spectral analyses with the LET in the 0.2-25 keV band down to separations of $\sim6-8$'', while the HET will offer relatively contamination free spectral analyses down to separations of $\sim10$'' (assuming apertures encompassing $\sim50\%$ EEF are employed). At small angular separations, this contamination can be mitigated by fitting LET and HET spectra simultaneously, and including model component constants that account for spectral contamination. 
\end{itemize}
HEX-P is poised to transform the landscape of dual AGN science, offering one of the cleanest methods of AGN detection that will be largely obscuration independent given the hard X-ray bandpass and enhanced faint source sensitivity of the HET. \textit{HEX-P} will play a key role in uncovering heretofore unconfirmed dual AGNs in candidate systems, determining the occupation fraction of hard X-ray emitting and obscured dual AGNs in known and currently unknown dual AGN populations, and studying the fueling and obscuration levels of dual AGNs across the merger sequence and as a function of AGN type in the local Universe. Such observations will refine previously proposed correlations between the projected pair separations of the AGNs and their X-ray luminosities \citep{koss2012,hou2020} and column densities \citep{guainazzi2021}. Bayesian detection algorithms akin to BAYMAX \citep{foord2019,foord2020,foord2021} will provide statistically powerful improvements to the detectability of closely-separated sources, and such Bayesian methods will provide more accurate constraints on the flux ratios of close angular pairs and thereby improve the statistical priors used during the spectral fitting process. Furthermore, with the high sensitivity and angular resolution of the LET, \textit{HEX-P} will allow the simultaneous examination of the host stellar populations via the detection of hot star formation-driven components in the soft X-rays, often observed in dual AGNs \citep[see Section~5.4 in ][]{pfeifle2023c}. Thus, \textit{HEX-P} will also probe connections between dual AGN activation and obscuration with the properties of the host stellar populations.

%\subsection{Figures}
%Frontiers requires figures to be submitted individually, in the same order as they are referred to in the manuscript. Figures will then be automatically embedded at the bottom of the submitted manuscript. Kindly ensure that each table and figure is mentioned in the text and in numerical order. Figures must be of sufficient resolution for publication. Figures which are not according to the guidelines will cause substantial delay during the production process. Please see \href{https://www.frontiersin.org/guidelines/author-guidelines#figure-and-table-guidelines}{here} for full figure guidelines. Cite figures with subfigures as figure \ref{fig:Subfigure 1} and \ref{fig:Subfigure 2}.

%\subsubsection{Permission to Reuse and Copyright}
%Figures, tables, and images will be published under a Creative Commons CC-BY licence and permission must be obtained for use of copyrighted material from other sources (including re-published/adapted/modified/partial figures and images from the internet). It is the responsibility of the authors to acquire the licenses, to follow any citation instructions requested by third-party rights holders, and cover any supplementary charges.
%%Figures, tables, and images will be published under a Creative Commons CC-BY licence and permission must be obtained for use of copyrighted material from other sources (including re-published/adapted/modified/partial figures and images from the internet). It is the responsibility of the authors to acquire the licenses, to follow any citation instructions requested by third-party rights holders, and cover any supplementary charges.

\section*{Conflict of Interest Statement}
%All financial, commercial or other relationships that might be perceived by the academic community as representing a potential conflict of interest must be disclosed. If no such relationship exists, authors will be asked to confirm the following statement: 

The authors declare that the research was conducted in the absence of any commercial or financial relationships that could be construed as a potential conflict of interest. R. W. P. is a member of the \textit{HEX-P} AGN Science Working Group and is also currently a co-deputy for the Line Emission Mapper (LEM) AGN Working Group; both of these are current mission concept studies.

\section*{Author Contributions}
R. W. P. led the work, developed the SOXS, SIXTE, and Jupyter Notebook scripts used to generate all of the simulated event files, science images, and spectroscopic data products for this study; he led all analysis of spatial resolution and spectral signal-to-noise. P. G. B. assisted with the generation of spectroscopic data products and performed the spectroscopic data analysis for the Arp 299 test case in this work. K. A. W. provided guidance for the scientific content of this work. J. B. developed the UltraNest-based Bayesian source detection algorithm used for the spatial resolution source detection analysis in this work. F. C. and P. G. B., as science lead and deputy lead for the \textit{HEX-P} AGN working group, provided general guidance for the science content of this work. All authors reviewed the manuscript and provided feedback to improve the state of the manuscript.

%The Author Contributions section is mandatory for all articles, including articles by sole authors. If an appropriate statement is not provided on submission, a standard one will be inserted during the production process. The Author Contributions statement must describe the contributions of individual authors referred to by their initials and, in doing so, all authors agree to be accountable for the content of the work. Please see  \href{https://www.frontiersin.org/guidelines/policies-and-publication-ethics#authorship-and-author-responsibilities}{here} for full authorship criteria.

\section*{Funding}
R.W.P. gratefully acknowledges support through an appointment to the NASA Postdoctoral Program at Goddard Space Flight Center, administered by ORAU through a contract with NASA.  The work of D.S. was carried out at the Jet Propulsion Laboratory, California Institute of Technology, under a contract with NASA. C.R. acknowledges support from Fondecyt Regular grant 1230345 and ANID BASAL project FB210003. E.K. acknowledges financial support from the Centre National d’Etudes Spatiales (CNES). L.M. is supported by the CITA National Fellowship.
%Details of all funding sources should be provided, including grant numbers if applicable. Please ensure to add all necessary funding information, as after publication this is no longer possible.

\section*{Acknowledgments}
We thank the SIXTE support team (with particular thanks to Ole König) for their assistance in generating spectra from simulated event files and for assistance with the documentation. %This is a short text to acknowledge the contributions of specific colleagues, institutions, or agencies that aided the efforts of the authors.

%\section*{Supplemental Data}
% \href{https://www.frontiersin.org/guidelines/author-guidelines#supplementary-material}{Supplementary Material} should be uploaded separately on submission, if there are Supplementary Figures, please include the caption in the same file as the figure. LaTeX Supplementary Material templates can be found in the Frontiers LaTeX folder.

\section*{Data Availability Statement}
The SIXTE and SOXS scripts along with the Jupyter Notebooks used to generate and analyze the simulated data sets for this study can be found in the following GitHub repository\footnote{\url{https://github.com/thatastroguy/HEXP}} owned by R. W. P. Note, the actual data products are not included in this remote repository due to the size of the files, but the scripts and Notebooks can be used to regenerate all of the data products created for this work. Copies of the simulated data are available upon request.
% Please see the availability of data guidelines for more information, at https://www.frontiersin.org/guidelines/policies-and-publication-ethics#materials-and-data-policies

\section{Appendix}

\begin{figure*}
    \includegraphics[width=0.99\linewidth]{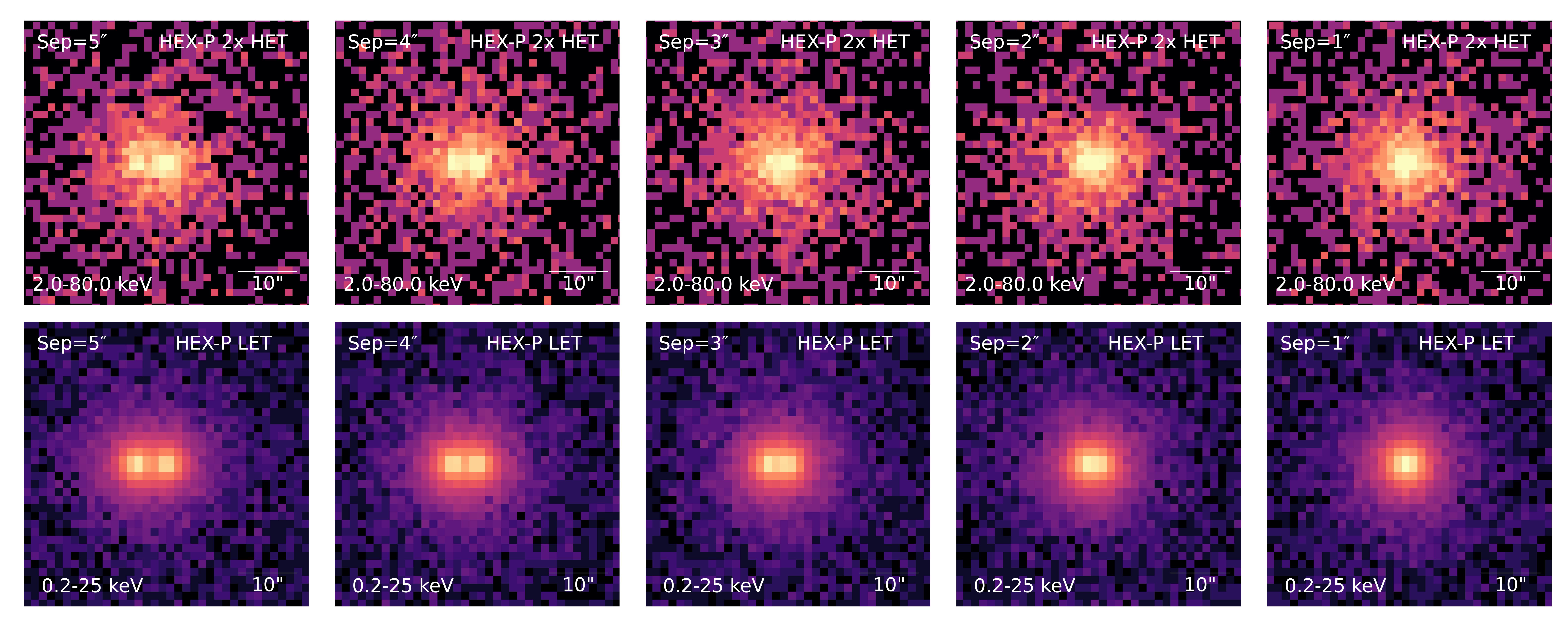}
    \caption{This 8-panel figure highlights the \textit{HEX-P} HET and LET imaging for pair separations ranging from 5'' down to 1''. With the exception of the choice of pair separations, these panels are identical in design as Figure~\ref{fig:LET} and \ref{fig:HET}. Here we can identify pairs of sources down to $\sim4''$ in the HET and $\sim3''$ in the LET by eye. See Section~\ref{sec:soudetect} on algorithmically detecting even more closely separated sources in \textit{HEX-P} imaging.}
    \label{fig:appendix_5as_to_1as}
\end{figure*}

\bibliographystyle{Frontiers-Harvard} %  Many Frontiers journals use the Harvard referencing system (Author-date), to find the style and resources for the journal you are submitting to: https://zendesk.frontiersin.org/hc/en-us/articles/360017860337-Frontiers-Reference-Styles-by-Journal. For Humanities and Social Sciences articles please include page numbers in the in-text citations 
\bibliography{test}

\begin{thebibliography}{113}
\providecommand{\natexlab}[1]{#1}
\expandafter\ifx\csname urlstyle\endcsname\relax
  \providecommand{\doi}[1]{doi:\discretionary{}{}{}#1}\else
  \providecommand{\doi}{doi:\discretionary{}{}{}\begingroup
  \urlstyle{rm}\Url}\fi
\providecommand{\selectlanguage}[1]{\relax}
\providecommand{\bibAnnoteFile}[1]{%
  \IfFileExists{#1}{\begin{quotation}\noindent\textsc{Key:} #1\\
  \textsc{Annotation:}\ \input{#1}\end{quotation}}{}}
\providecommand{\bibAnnote}[2]{%
  \begin{quotation}\noindent\textsc{Key:} #1\\
  \textsc{Annotation:}\ #2\end{quotation}}

\bibitem[{{Akaike}(1974)}]{akaike1974}
{Akaike}, H. (1974).
\newblock {A New Look at the Statistical Model Identification}.
\newblock \emph{IEEE Transactions on Automatic Control} 19, 716--723
\bibAnnoteFile{akaike1974}

\bibitem[{{Alonso-Herrero} et~al.(2013){Alonso-Herrero}, {Roche}, {Esquej},
  {Gonz{\'a}lez-Mart{\'\i}n}, {Pereira-Santaella}, {Ramos Almeida}
  et~al.}]{alonsoherrero2013}
{Alonso-Herrero}, A., {Roche}, P.~F., {Esquej}, P., {Gonz{\'a}lez-Mart{\'\i}n},
  O., {Pereira-Santaella}, M., {Ramos Almeida}, C., et~al. (2013).
\newblock {Uncovering the Deeply Embedded Active Galactic Nucleus Activity in
  the Nuclear Regions of the Interacting Galaxy Arp 299}.
\newblock \emph{\apjl} 779, L14.
\newblock \doi{10.1088/2041-8205/779/1/L14}
\bibAnnoteFile{alonsoherrero2013}

\bibitem[{{Amaro-Seoane} et~al.(2023){Amaro-Seoane}, {Andrews}, {Arca Sedda},
  {Askar}, {Baghi}, {Balasov} et~al.}]{amaro-seoane2023}
{Amaro-Seoane}, P., {Andrews}, J., {Arca Sedda}, M., {Askar}, A., {Baghi}, Q.,
  {Balasov}, R., et~al. (2023).
\newblock {Astrophysics with the Laser Interferometer Space Antenna}.
\newblock \emph{Living Reviews in Relativity} 26, 2.
\newblock \doi{10.1007/s41114-022-00041-y}
\bibAnnoteFile{amaro-seoane2023}

\bibitem[{{Arnaud}(1996)}]{Arnaud1996}
{Arnaud}, K.~A. (1996).
\newblock {XSPEC: The First Ten Years}.
\newblock In \emph{Astronomical Data Analysis Software and Systems V}, eds.
  G.~H. {Jacoby} and J.~{Barnes}. vol. 101 of \emph{Astronomical Society of the
  Pacific Conference Series}, 17
\bibAnnoteFile{Arnaud1996}

\bibitem[{{Ballo} et~al.(2004){Ballo}, {Braito}, {Della Ceca}, {Maraschi},
  {Tavecchio}, and {Dadina}}]{ballo2004}
{Ballo}, L., {Braito}, V., {Della Ceca}, R., {Maraschi}, L., {Tavecchio}, F.,
  and {Dadina}, M. (2004).
\newblock {Arp 299: A Second Merging System with Two Active Nuclei?}
\newblock \emph{\apj} 600, 634--639.
\newblock \doi{10.1086/379887}
\bibAnnoteFile{ballo2004}

\bibitem[{{Balokovi{\'c}} et~al.(2018){Balokovi{\'c}}, {Brightman}, {Harrison},
  {Comastri}, {Ricci}, {Buchner} et~al.}]{balokovic2018}
{Balokovi{\'c}}, M., {Brightman}, M., {Harrison}, F.~A., {Comastri}, A.,
  {Ricci}, C., {Buchner}, J., et~al. (2018).
\newblock {New Spectral Model for Constraining Torus Covering Factors from
  Broadband X-Ray Spectra of Active Galactic Nuclei}.
\newblock \emph{\apj} 854, 42.
\newblock \doi{10.3847/1538-4357/aaa7eb}
\bibAnnoteFile{balokovic2018}

\bibitem[{{Barnes} and {Hernquist}(1996)}]{barnes1996}
{Barnes}, J.~E. and {Hernquist}, L. (1996).
\newblock {Transformations of Galaxies. II. Gasdynamics in Merging Disk
  Galaxies}.
\newblock \emph{\apj} 471, 115.
\newblock \doi{10.1086/177957}
\bibAnnoteFile{barnes1996}

\bibitem[{{Barnes} and {Hernquist}(1991)}]{barnes1991}
{Barnes}, J.~E. and {Hernquist}, L.~E. (1991).
\newblock {Fueling Starburst Galaxies with Gas-rich Mergers}.
\newblock \emph{\apjl} 370, L65.
\newblock \doi{10.1086/185978}
\bibAnnoteFile{barnes1991}

\bibitem[{{Barrows} et~al.(2023){Barrows}, {Comerford}, {Stern}, and
  {Assef}}]{barrows2023}
{Barrows}, R.~S., {Comerford}, J.~M., {Stern}, D., and {Assef}, R.~J. (2023).
\newblock {A Census of WISE-selected Dual and Offset AGNs Across the Sky: New
  Constraints on Merger-driven Triggering of Obscured AGNs}.
\newblock \emph{\apj} 951, 92.
\newblock \doi{10.3847/1538-4357/acd2d3}
\bibAnnoteFile{barrows2023}

\bibitem[{{Bianchi} et~al.(2008){Bianchi}, {Chiaberge}, {Piconcelli},
  {Guainazzi}, and {Matt}}]{bianchi2008}
{Bianchi}, S., {Chiaberge}, M., {Piconcelli}, E., {Guainazzi}, M., and {Matt},
  G. (2008).
\newblock {Chandra unveils a binary active galactic nucleus in Mrk 463}.
\newblock \emph{\mnras} 386, 105--110.
\newblock \doi{10.1111/j.1365-2966.2008.13078.x}
\bibAnnoteFile{bianchi2008}

\bibitem[{{Blecha} et~al.(2018){Blecha}, {Snyder}, {Satyapal}, and
  {Ellison}}]{blecha2018}
{Blecha}, L., {Snyder}, G.~F., {Satyapal}, S., and {Ellison}, S.~L. (2018).
\newblock {The power of infrared AGN selection in mergers: a theoretical
  study}.
\newblock \emph{\mnras} 478, 3056--3071.
\newblock \doi{10.1093/mnras/sty1274}
\bibAnnoteFile{blecha2018}

\bibitem[{{Buchner}(2019)}]{Buchner2019}
{Buchner}, J. (2019).
\newblock {Collaborative Nested Sampling: Big Data versus Complex Physical
  Models}.
\newblock \emph{\pasp} 131, 108005.
\newblock \doi{10.1088/1538-3873/aae7fc}
\bibAnnoteFile{Buchner2019}

\bibitem[{{Buchner}(2021{\natexlab{a}})}]{BXAsoftwarepaper}
{Buchner}, J. (2021{\natexlab{a}}).
\newblock {Bayesian X-ray Analysis (BXA) v4.0}.
\newblock \emph{The Journal of Open Source Software} 6, 3045.
\newblock \doi{10.21105/joss.03045}
\bibAnnoteFile{BXAsoftwarepaper}

\bibitem[{{Buchner}(2021{\natexlab{b}})}]{UltraNest2021JOSS}
{Buchner}, J. (2021{\natexlab{b}}).
\newblock {UltraNest - a robust, general purpose Bayesian inference engine}.
\newblock \emph{The Journal of Open Source Software} 6, 3001.
\newblock \doi{10.21105/joss.03001}
\bibAnnoteFile{UltraNest2021JOSS}

\bibitem[{Buchner(2023)}]{Buchner2021c}
Buchner, J. (2023).
\newblock {Nested sampling methods}.
\newblock \emph{Statistics Surveys} 17, arXiv:2101.09675.
\newblock \doi{10.1214/23-SS144}
\bibAnnoteFile{Buchner2021c}

\bibitem[{{Buchner} et~al.(2014{\natexlab{a}}){Buchner}, {Georgakakis},
  {Nandra}, {Hsu}, {Rangel}, {Brightman} et~al.}]{Buchner14}
{Buchner}, J., {Georgakakis}, A., {Nandra}, K., {Hsu}, L., {Rangel}, C.,
  {Brightman}, M., et~al. (2014{\natexlab{a}}).
\newblock {X-ray spectral modelling of the AGN obscuring region in the CDFS:
  Bayesian model selection and catalogue}.
\newblock \emph{\aap} 564, A125.
\newblock \doi{10.1051/0004-6361/201322971}
\bibAnnoteFile{Buchner14}

\bibitem[{{Buchner} et~al.(2014{\natexlab{b}}){Buchner}, {Georgakakis},
  {Nandra}, {Hsu}, {Rangel}, {Brightman} et~al.}]{buchner2014}
{Buchner}, J., {Georgakakis}, A., {Nandra}, K., {Hsu}, L., {Rangel}, C.,
  {Brightman}, M., et~al. (2014{\natexlab{b}}).
\newblock {X-ray spectral modelling of the AGN obscuring region in the CDFS:
  Bayesian model selection and catalogue}.
\newblock \emph{\aap} 564, A125.
\newblock \doi{10.1051/0004-6361/201322971}
\bibAnnoteFile{buchner2014}

\bibitem[{{Burke-Spolaor} et~al.(2018){Burke-Spolaor}, {Blecha}, {Bogdanovic},
  {Comerford}, {Lazio}, {Liu} et~al.}]{burkespolaor2018}
{Burke-Spolaor}, S., {Blecha}, L., {Bogdanovic}, T., {Comerford}, J.~M.,
  {Lazio}, T. J.~W., {Liu}, X., et~al. (2018).
\newblock {The Next-Generation Very Large Array: Supermassive Black Hole Pairs
  and Binaries}.
\newblock \emph{arXiv e-prints} ,
  arXiv:1808.04368\doi{10.48550/arXiv.1808.04368}
\bibAnnoteFile{burkespolaor2018}

\bibitem[{{Callegari} et~al.(2011){Callegari}, {Kazantzidis}, {Mayer}, {Colpi},
  {Bellovary}, {Quinn} et~al.}]{callegari2011}
{Callegari}, S., {Kazantzidis}, S., {Mayer}, L., {Colpi}, M., {Bellovary},
  J.~M., {Quinn}, T., et~al. (2011).
\newblock {Growing Massive Black Hole Pairs in Minor Mergers of Disk Galaxies}.
\newblock \emph{\apj} 729, 85.
\newblock \doi{10.1088/0004-637X/729/2/85}
\bibAnnoteFile{callegari2011}

\bibitem[{{Callegari} et~al.(2009){Callegari}, {Mayer}, {Kazantzidis}, {Colpi},
  {Governato}, {Quinn} et~al.}]{callegari2009}
{Callegari}, S., {Mayer}, L., {Kazantzidis}, S., {Colpi}, M., {Governato}, F.,
  {Quinn}, T., et~al. (2009).
\newblock {Pairing of Supermassive Black Holes in Unequal-Mass Galaxy Mergers}.
\newblock \emph{\apjl} 696, L89--L92.
\newblock \doi{10.1088/0004-637X/696/1/L89}
\bibAnnoteFile{callegari2009}

\bibitem[{{Capelo} et~al.(2017){Capelo}, {Dotti}, {Volonteri}, {Mayer},
  {Bellovary}, and {Shen}}]{capelo2017}
{Capelo}, P.~R., {Dotti}, M., {Volonteri}, M., {Mayer}, L., {Bellovary}, J.~M.,
  and {Shen}, S. (2017).
\newblock {A survey of dual active galactic nuclei in simulations of galaxy
  mergers: frequency and properties}.
\newblock \emph{\mnras} 469, 4437--4454.
\newblock \doi{10.1093/mnras/stx1067}
\bibAnnoteFile{capelo2017}

\bibitem[{{Capelo} et~al.(2015){Capelo}, {Volonteri}, {Dotti}, {Bellovary},
  {Mayer}, and {Governato}}]{capelo2015}
{Capelo}, P.~R., {Volonteri}, M., {Dotti}, M., {Bellovary}, J.~M., {Mayer}, L.,
  and {Governato}, F. (2015).
\newblock {Growth and activity of black holes in galaxy mergers with varying
  mass ratios}.
\newblock \emph{\mnras} 447, 2123--2143.
\newblock \doi{10.1093/mnras/stu2500}
\bibAnnoteFile{capelo2015}

\bibitem[{{Ciurlo} et~al.(2023){Ciurlo}, {Mannucci}, {Yeh}, {Amiri},
  {Carniani}, {Cicone} et~al.}]{ciurlo2023}
{Ciurlo}, A., {Mannucci}, F., {Yeh}, S., {Amiri}, A., {Carniani}, S., {Cicone},
  C., et~al. (2023).
\newblock {New multiple AGN systems with subarcsec separation: Confirmation of
  candidates selected via the novel GMP method}.
\newblock \emph{\aap} 671, L4.
\newblock \doi{10.1051/0004-6361/202345853}
\bibAnnoteFile{ciurlo2023}

\bibitem[{{Civano} et~al.(2012){Civano}, {Elvis}, {Lanzuisi}, {Aldcroft},
  {Trichas}, {Bongiorno} et~al.}]{civano2012}
{Civano}, F., {Elvis}, M., {Lanzuisi}, G., {Aldcroft}, T., {Trichas}, M.,
  {Bongiorno}, A., et~al. (2012).
\newblock {Chandra High-resolution observations of CID-42, a Candidate
  Recoiling Supermassive Black Hole}.
\newblock \emph{\apj} 752, 49.
\newblock \doi{10.1088/0004-637X/752/1/49}
\bibAnnoteFile{civano2012}

\bibitem[{{Civano} et~al.(2010){Civano}, {Elvis}, {Lanzuisi}, {Jahnke},
  {Zamorani}, {Blecha} et~al.}]{civano2010}
{Civano}, F., {Elvis}, M., {Lanzuisi}, G., {Jahnke}, K., {Zamorani}, G.,
  {Blecha}, L., et~al. (2010).
\newblock {A Runaway Black Hole in COSMOS: Gravitational Wave or Slingshot
  Recoil?}
\newblock \emph{\apj} 717, 209--222.
\newblock \doi{10.1088/0004-637X/717/1/209}
\bibAnnoteFile{civano2010}

\bibitem[{{Civano} et~al.(2015){Civano}, {Hickox}, {Puccetti}, {Comastri},
  {Mullaney}, {Zappacosta} et~al.}]{civano2015}
{Civano}, F., {Hickox}, R.~C., {Puccetti}, S., {Comastri}, A., {Mullaney},
  J.~R., {Zappacosta}, L., et~al. (2015).
\newblock {The Nustar Extragalactic Surveys: Overview and Catalog from the
  COSMOS Field}.
\newblock \emph{\apj} 808, 185.
\newblock \doi{10.1088/0004-637X/808/2/185}
\bibAnnoteFile{civano2015}

\bibitem[{{Comerford} et~al.(2012){Comerford}, {Gerke}, {Stern}, {Cooper},
  {Weiner}, {Newman} et~al.}]{comerford2012}
{Comerford}, J.~M., {Gerke}, B.~F., {Stern}, D., {Cooper}, M.~C., {Weiner},
  B.~J., {Newman}, J.~A., et~al. (2012).
\newblock {Kiloparsec-scale Spatial Offsets in Double-peaked Narrow-line Active
  Galactic Nuclei. I. Markers for Selection of Compelling Dual Active Galactic
  Nucleus Candidates}.
\newblock \emph{\apj} 753, 42.
\newblock \doi{10.1088/0004-637X/753/1/42}
\bibAnnoteFile{comerford2012}

\bibitem[{{Comerford} et~al.(2009){Comerford}, {Griffith}, {Gerke}, {Cooper},
  {Newman}, {Davis} et~al.}]{comerford2009}
{Comerford}, J.~M., {Griffith}, R.~L., {Gerke}, B.~F., {Cooper}, M.~C.,
  {Newman}, J.~A., {Davis}, M., et~al. (2009).
\newblock {1.75 h $^{-1}$ kpc Separation Dual Active Galactic Nuclei at z =
  0.36 in the Cosmos Field}.
\newblock \emph{\apjl} 702, L82--L86.
\newblock \doi{10.1088/0004-637X/702/1/L82}
\bibAnnoteFile{comerford2009}

\bibitem[{{Comerford} et~al.(2015){Comerford}, {Pooley}, {Barrows}, {Greene},
  {Zakamska}, {Madejski} et~al.}]{comerford2015}
{Comerford}, J.~M., {Pooley}, D., {Barrows}, R.~S., {Greene}, J.~E.,
  {Zakamska}, N.~L., {Madejski}, G.~M., et~al. (2015).
\newblock {Merger-driven Fueling of Active Galactic Nuclei: Six Dual and Offset
  AGNs Discovered with Chandra and Hubble Space Telescope Observations}.
\newblock \emph{\apj} 806, 219.
\newblock \doi{10.1088/0004-637X/806/2/219}
\bibAnnoteFile{comerford2015}

\bibitem[{{Comerford} et~al.(2011){Comerford}, {Pooley}, {Gerke}, and
  {Madejski}}]{comerford2011}
{Comerford}, J.~M., {Pooley}, D., {Gerke}, B.~F., and {Madejski}, G.~M. (2011).
\newblock {Chandra Observations of a 1.9 kpc Separation Double X-Ray Source in
  a Candidate Dual Active Galactic Nucleus Galaxy at z = 0.16}.
\newblock \emph{\apjl} 737, L19.
\newblock \doi{10.1088/2041-8205/737/1/L19}
\bibAnnoteFile{comerford2011}

\bibitem[{{Dauser} et~al.(2019){Dauser}, {Falkner}, {Lorenz}, {Kirsch},
  {Peille}, {Cucchetti} et~al.}]{dauser2019}
{Dauser}, T., {Falkner}, S., {Lorenz}, M., {Kirsch}, C., {Peille}, P.,
  {Cucchetti}, E., et~al. (2019).
\newblock {SIXTE: a generic X-ray instrument simulation toolkit}.
\newblock \emph{\aap} 630, A66.
\newblock \doi{10.1051/0004-6361/201935978}
\bibAnnoteFile{dauser2019}

\bibitem[{{De Rosa} et~al.(2018){De Rosa}, {Vignali}, {Husemann}, {Bianchi},
  {Bogdanovi{\'c}}, {Guainazzi} et~al.}]{derosa2018}
{De Rosa}, A., {Vignali}, C., {Husemann}, B., {Bianchi}, S., {Bogdanovi{\'c}},
  T., {Guainazzi}, M., et~al. (2018).
\newblock {Disclosing the properties of low-redshift dual AGN through
  XMM-Newton and SDSS spectroscopy}.
\newblock \emph{\mnras} 480, 1639--1655.
\newblock \doi{10.1093/mnras/sty1867}
\bibAnnoteFile{derosa2018}

\bibitem[{{De Rosa} et~al.(2023){De Rosa}, {Vignali}, {Severgnini}, {Bianchi},
  {Bogdanovi{\'c}}, {Charisi} et~al.}]{derosa2023}
{De Rosa}, A., {Vignali}, C., {Severgnini}, P., {Bianchi}, S.,
  {Bogdanovi{\'c}}, T., {Charisi}, M., et~al. (2023).
\newblock {The X-ray view of optically selected dual AGN}.
\newblock \emph{\mnras} 519, 5149--5160.
\newblock \doi{10.1093/mnras/stac3664}
\bibAnnoteFile{derosa2023}

\bibitem[{{Ellison} et~al.(2011){Ellison}, {Patton}, {Mendel}, and
  {Scudder}}]{ellison2011}
{Ellison}, S.~L., {Patton}, D.~R., {Mendel}, J.~T., and {Scudder}, J.~M.
  (2011).
\newblock {Galaxy pairs in the Sloan Digital Sky Survey - IV. Interactions
  trigger active galactic nuclei}.
\newblock \emph{\mnras} 418, 2043--2053.
\newblock \doi{10.1111/j.1365-2966.2011.19624.x}
\bibAnnoteFile{ellison2011}

\bibitem[{{Ellison} et~al.(2017){Ellison}, {Secrest}, {Mendel}, {Satyapal}, and
  {Simard}}]{ellison2017}
{Ellison}, S.~L., {Secrest}, N.~J., {Mendel}, J.~T., {Satyapal}, S., and
  {Simard}, L. (2017).
\newblock {Discovery of a dual active galactic nucleus with
  {\ensuremath{\sim}}8 kpc separation}.
\newblock \emph{\mnras} 470, L49--L53.
\newblock \doi{10.1093/mnrasl/slx076}
\bibAnnoteFile{ellison2017}

\bibitem[{{Eraerds} et~al.(2021){Eraerds}, {Antonelli}, {Davis}, {Hall},
  {Hetherington}, {Holland} et~al.}]{Eraerds2021}
{Eraerds}, T., {Antonelli}, V., {Davis}, C., {Hall}, D., {Hetherington}, O.,
  {Holland}, A., et~al. (2021).
\newblock {Enhanced simulations on the Athena/Wide Field Imager instrumental
  background}.
\newblock \emph{Journal of Astronomical Telescopes, Instruments, and Systems}
  7, 034001.
\newblock \doi{10.1117/1.JATIS.7.3.034001}
\bibAnnoteFile{Eraerds2021}

\bibitem[{{Foord} et~al.(2020){Foord}, {G{\"u}ltekin}, {Nevin}, {Comerford},
  {Hodges-Kluck}, {Barrows} et~al.}]{foord2020}
{Foord}, A., {G{\"u}ltekin}, K., {Nevin}, R., {Comerford}, J.~M.,
  {Hodges-Kluck}, E., {Barrows}, R.~S., et~al. (2020).
\newblock {A Second Look at 12 Candidate Dual AGNs Using BAYMAX}.
\newblock \emph{\apj} 892, 29.
\newblock \doi{10.3847/1538-4357/ab72fa}
\bibAnnoteFile{foord2020}

\bibitem[{{Foord} et~al.(2019){Foord}, {G{\"u}ltekin}, {Reynolds},
  {Hodges-Kluck}, {Cackett}, {Comerford} et~al.}]{foord2019}
{Foord}, A., {G{\"u}ltekin}, K., {Reynolds}, M.~T., {Hodges-Kluck}, E.,
  {Cackett}, E.~M., {Comerford}, J.~M., et~al. (2019).
\newblock {A Bayesian Analysis of SDSS J0914+0853, a Low-mass Dual AGN
  Candidate}.
\newblock \emph{\apj} 877, 17.
\newblock \doi{10.3847/1538-4357/ab18a3}
\bibAnnoteFile{foord2019}

\bibitem[{{Foord} et~al.(2021){Foord}, {G{\"u}ltekin}, {Runnoe}, and
  {Koss}}]{foord2021}
{Foord}, A., {G{\"u}ltekin}, K., {Runnoe}, J.~C., and {Koss}, M.~J. (2021).
\newblock {AGN Triality of Triple Mergers: Detection of Faint X-Ray Point
  Sources}.
\newblock \emph{\apj} 907, 71.
\newblock \doi{10.3847/1538-4357/abce5d}
\bibAnnoteFile{foord2021}

\bibitem[{{Frey} et~al.(2012){Frey}, {Paragi}, {An}, and
  {Gab{\'a}nyi}}]{frey2012}
{Frey}, S., {Paragi}, Z., {An}, T., and {Gab{\'a}nyi}, K.~{\'E}. (2012).
\newblock {Two in one? A possible dual radio-emitting nucleus in the quasar
  SDSS J1425+3231}.
\newblock \emph{\mnras} 425, 1185--1191.
\newblock \doi{10.1111/j.1365-2966.2012.21491.x}
\bibAnnoteFile{frey2012}

\bibitem[{{Fu} et~al.(2015{\natexlab{a}}){Fu}, {Myers}, {Djorgovski}, {Yan},
  {Wrobel}, and {Stockton}}]{fu2015a}
{Fu}, H., {Myers}, A.~D., {Djorgovski}, S.~G., {Yan}, L., {Wrobel}, J.~M., and
  {Stockton}, A. (2015{\natexlab{a}}).
\newblock {Radio-selected Binary Active Galactic Nuclei from the Very Large
  Array Stripe 82 Survey}.
\newblock \emph{\apj} 799, 72.
\newblock \doi{10.1088/0004-637X/799/1/72}
\bibAnnoteFile{fu2015a}

\bibitem[{{Fu} et~al.(2015{\natexlab{b}}){Fu}, {Wrobel}, {Myers}, {Djorgovski},
  and {Yan}}]{fu2015b}
{Fu}, H., {Wrobel}, J.~M., {Myers}, A.~D., {Djorgovski}, S.~G., and {Yan}, L.
  (2015{\natexlab{b}}).
\newblock {Binary Active Galactic Nuclei in Stripe 82: Constraints on
  Synchronized Black Hole Accretion in Major Mergers}.
\newblock \emph{\apjl} 815, L6.
\newblock \doi{10.1088/2041-8205/815/1/L6}
\bibAnnoteFile{fu2015b}

\bibitem[{{Fu} et~al.(2011){Fu}, {Zhang}, {Assef}, {Stockton}, {Myers}, {Yan}
  et~al.}]{fu2011}
{Fu}, H., {Zhang}, Z.-Y., {Assef}, R.~J., {Stockton}, A., {Myers}, A.~D.,
  {Yan}, L., et~al. (2011).
\newblock {A Kiloparsec-scale Binary Active Galactic Nucleus Confirmed by the
  Expanded Very Large Array}.
\newblock \emph{\apjl} 740, L44.
\newblock \doi{10.1088/2041-8205/740/2/L44}
\bibAnnoteFile{fu2011}

\bibitem[{{Gab{\'a}nyi} et~al.(2016){Gab{\'a}nyi}, {An}, {Frey}, {Komossa},
  {Paragi}, {Hong} et~al.}]{gabanyi2016}
{Gab{\'a}nyi}, K.~{\'E}., {An}, T., {Frey}, S., {Komossa}, S., {Paragi}, Z.,
  {Hong}, X.~Y., et~al. (2016).
\newblock {Four Dual AGN Candidates Observed with the VLBA}.
\newblock \emph{\apj} 826, 106.
\newblock \doi{10.3847/0004-637X/826/2/106}
\bibAnnoteFile{gabanyi2016}

\bibitem[{{Gab{\'a}nyi} et~al.(2014){Gab{\'a}nyi}, {Frey}, {Xiao}, {Paragi},
  {An}, {Kun} et~al.}]{gabanyi2014}
{Gab{\'a}nyi}, K.~{\'E}., {Frey}, S., {Xiao}, T., {Paragi}, Z., {An}, T.,
  {Kun}, E., et~al. (2014).
\newblock {A single radio-emitting nucleus in the dual AGN candidate NGC 5515}.
\newblock \emph{\mnras} 443, 1509--1514.
\newblock \doi{10.1093/mnras/stu1234}
\bibAnnoteFile{gabanyi2014}

\bibitem[{{Ge} et~al.(2012){Ge}, {Hu}, {Wang}, {Bai}, and {Zhang}}]{ge2012}
{Ge}, J.-Q., {Hu}, C., {Wang}, J.-M., {Bai}, J.-M., and {Zhang}, S. (2012).
\newblock {Double-peaked Narrow Emission-line Galaxies from the Sloan Digital
  Sky Survey. I. Sample and Basic Properties}.
\newblock \emph{\apjs} 201, 31.
\newblock \doi{10.1088/0067-0049/201/2/31}
\bibAnnoteFile{ge2012}

\bibitem[{{Goulding} et~al.(2019){Goulding}, {Pardo}, {Greene}, {Mingarelli},
  {Nyland}, and {Strauss}}]{goulding2019}
{Goulding}, A.~D., {Pardo}, K., {Greene}, J.~E., {Mingarelli}, C. M.~F.,
  {Nyland}, K., and {Strauss}, M.~A. (2019).
\newblock {Discovery of a Close-separation Binary Quasar at the Heart of a z
  {\ensuremath{\sim}} 0.2 Merging Galaxy and Its Implications for Low-frequency
  Gravitational Waves}.
\newblock \emph{\apjl} 879, L21.
\newblock \doi{10.3847/2041-8213/ab2a14}
\bibAnnoteFile{goulding2019}

\bibitem[{{Gross} et~al.(2019){Gross}, {Fu}, {Myers}, {Wrobel}, and
  {Djorgovski}}]{gross2019}
{Gross}, A.~C., {Fu}, H., {Myers}, A.~D., {Wrobel}, J.~M., and {Djorgovski},
  S.~G. (2019).
\newblock {X-Ray Properties of Radio-selected Dual Active Galactic Nuclei}.
\newblock \emph{\apj} 883, 50.
\newblock \doi{10.3847/1538-4357/ab3795}
\bibAnnoteFile{gross2019}

\bibitem[{{Guainazzi} et~al.(2021){Guainazzi}, {De Rosa}, {Bianchi},
  {Husemann}, {Bogdanovic}, {Komossa} et~al.}]{guainazzi2021}
{Guainazzi}, M., {De Rosa}, A., {Bianchi}, S., {Husemann}, B., {Bogdanovic},
  T., {Komossa}, S., et~al. (2021).
\newblock {An XMM-Newton study of active-inactive galaxy pairs}.
\newblock \emph{\mnras} 504, 393--405.
\newblock \doi{10.1093/mnras/stab808}
\bibAnnoteFile{guainazzi2021}

\bibitem[{{Guainazzi} et~al.(2005){Guainazzi}, {Piconcelli},
  {Jim{\'e}nez-Bail{\'o}n}, and {Matt}}]{guainazzi2005}
{Guainazzi}, M., {Piconcelli}, E., {Jim{\'e}nez-Bail{\'o}n}, E., and {Matt}, G.
  (2005).
\newblock {The early stage of a cosmic collision? XMM-Newton unveils two
  obscured AGN in the galaxy pair ESO509-IG066}.
\newblock \emph{\aap} 429, L9--L12.
\newblock \doi{10.1051/0004-6361:200400104}
\bibAnnoteFile{guainazzi2005}

\bibitem[{{Harrison} et~al.(2013){Harrison}, {Craig}, {Christensen}, {Hailey},
  {Zhang}, {Boggs} et~al.}]{harrison2013}
{Harrison}, F.~A., {Craig}, W.~W., {Christensen}, F.~E., {Hailey}, C.~J.,
  {Zhang}, W.~W., {Boggs}, S.~E., et~al. (2013).
\newblock {The Nuclear Spectroscopic Telescope Array (NuSTAR) High-energy X-Ray
  Mission}.
\newblock \emph{\apj} 770, 103.
\newblock \doi{10.1088/0004-637X/770/2/103}
\bibAnnoteFile{harrison2013}

\bibitem[{{Hopkins} et~al.(2006){Hopkins}, {Hernquist}, {Cox}, {Di Matteo},
  {Robertson}, and {Springel}}]{hopkins2006}
{Hopkins}, P.~F., {Hernquist}, L., {Cox}, T.~J., {Di Matteo}, T., {Robertson},
  B., and {Springel}, V. (2006).
\newblock {A Unified, Merger-driven Model of the Origin of Starbursts, Quasars,
  the Cosmic X-Ray Background, Supermassive Black Holes, and Galaxy Spheroids}.
\newblock \emph{\apjs} 163, 1--49.
\newblock \doi{10.1086/499298}
\bibAnnoteFile{hopkins2006}

\bibitem[{{Hopkins} et~al.(2008){Hopkins}, {Hernquist}, {Cox}, and
  {Kere{\v{s}}}}]{hopkins2008}
{Hopkins}, P.~F., {Hernquist}, L., {Cox}, T.~J., and {Kere{\v{s}}}, D. (2008).
\newblock {A Cosmological Framework for the Co-Evolution of Quasars,
  Supermassive Black Holes, and Elliptical Galaxies. I. Galaxy Mergers and
  Quasar Activity}.
\newblock \emph{\apjs} 175, 356--389.
\newblock \doi{10.1086/524362}
\bibAnnoteFile{hopkins2008}

\bibitem[{{Hou} et~al.(2020){Hou}, {Li}, and {Liu}}]{hou2020}
{Hou}, M., {Li}, Z., and {Liu}, X. (2020).
\newblock {A Chandra X-Ray Survey of Optically Selected AGN Pairs}.
\newblock \emph{\apj} 900, 79.
\newblock \doi{10.3847/1538-4357/aba4a7}
\bibAnnoteFile{hou2020}

\bibitem[{{Hou} et~al.(2023){Hou}, {Li}, {Liu}, {Li}, {Li}, {Wang}
  et~al.}]{hou2023}
{Hou}, M., {Li}, Z., {Liu}, X., {Li}, Z., {Li}, R., {Wang}, R., et~al. (2023).
\newblock {NOEMA Detection of Circumnuclear Molecular Gas in X-Ray Weak Dual
  Active Galactic Nuclei: No Evidence for Heavy Obscuration}.
\newblock \emph{\apj} 943, 50.
\newblock \doi{10.3847/1538-4357/acaaf9}
\bibAnnoteFile{hou2023}

\bibitem[{{Hou} et~al.(2019){Hou}, {Liu}, {Guo}, {Li}, {Shen}, and
  {Green}}]{hou2019}
{Hou}, M., {Liu}, X., {Guo}, H., {Li}, Z., {Shen}, Y., and {Green}, P.~J.
  (2019).
\newblock {Active Galactic Nucleus Pairs from the Sloan Digital Sky Survey.
  III. Chandra X-Ray Observations Unveil Obscured Double Nuclei}.
\newblock \emph{\apj} 882, 41.
\newblock \doi{10.3847/1538-4357/ab3225}
\bibAnnoteFile{hou2019}

\bibitem[{{Hwang} et~al.(2020){Hwang}, {Hamer}, {Zakamska}, and
  {Schlaufman}}]{hwang2020}
{Hwang}, H.-C., {Hamer}, J.~H., {Zakamska}, N.~L., and {Schlaufman}, K.~C.
  (2020).
\newblock {Very wide companion fraction from Gaia DR2: A weak or no enhancement
  for hot Jupiter hosts, and a strong enhancement for contact binaries}.
\newblock \emph{\mnras} 497, 2250--2259.
\newblock \doi{10.1093/mnras/staa2124}
\bibAnnoteFile{hwang2020}

\bibitem[{{Imanishi} et~al.(2020){Imanishi}, {Kawamuro}, {Kikuta}, {Nakano},
  and {Saito}}]{imanishi2020}
{Imanishi}, M., {Kawamuro}, T., {Kikuta}, S., {Nakano}, S., and {Saito}, Y.
  (2020).
\newblock {Subaru Infrared Adaptive Optics-assisted High-spatial-resolution
  Imaging Search for Luminous Dual Active Galactic Nuclei in Nearby
  Ultraluminous Infrared Galaxies}.
\newblock \emph{\apj} 891, 140.
\newblock \doi{10.3847/1538-4357/ab733e}
\bibAnnoteFile{imanishi2020}

\bibitem[{{Imanishi} and {Saito}(2014)}]{imanishi2014}
{Imanishi}, M. and {Saito}, Y. (2014).
\newblock {Subaru Adaptive-optics High-spatial-resolution Infrared K- and
  L'-band Imaging Search for Deeply Buried Dual AGNs in Merging Galaxies}.
\newblock \emph{\apj} 780, 106.
\newblock \doi{10.1088/0004-637X/780/1/106}
\bibAnnoteFile{imanishi2014}

\bibitem[{{Ivezi{\'c}} et~al.(2019){Ivezi{\'c}}, {Kahn}, {Tyson}, {Abel},
  {Acosta}, {Allsman} et~al.}]{ivezic2019}
{Ivezi{\'c}}, {\v Z}., {Kahn}, S.~M., {Tyson}, J.~A., {Abel}, B., {Acosta}, E.,
  {Allsman}, R., et~al. (2019).
\newblock {LSST: From Science Drivers to Reference Design and Anticipated Data
  Products}.
\newblock \emph{\apj} 873, 111.
\newblock \doi{10.3847/1538-4357/ab042c}
\bibAnnoteFile{ivezic2019}

\bibitem[{{Iwasawa} et~al.(2020){Iwasawa}, {Ricci}, {Privon},
  {Torres-Alb{\`a}}, {Inami}, {Charmandaris} et~al.}]{iwasawa2020}
{Iwasawa}, K., {Ricci}, C., {Privon}, G.~C., {Torres-Alb{\`a}}, N., {Inami},
  H., {Charmandaris}, V., et~al. (2020).
\newblock {A Compton-thick nucleus in the dual active galactic nuclei of Mrk
  266}.
\newblock \emph{\aap} 640, A95.
\newblock \doi{10.1051/0004-6361/202038513}
\bibAnnoteFile{iwasawa2020}

\bibitem[{{Iwasawa} et~al.(2011){Iwasawa}, {Sanders}, {Teng}, {U}, {Armus},
  {Evans} et~al.}]{iwasawa2011}
{Iwasawa}, K., {Sanders}, D.~B., {Teng}, S.~H., {U}, V., {Armus}, L., {Evans},
  A.~S., et~al. (2011).
\newblock {C-GOALS: Chandra observations of a complete sample of luminous
  infrared galaxies from the IRAS Revised Bright Galaxy Survey}.
\newblock \emph{\aap} 529, A106.
\newblock \doi{10.1051/0004-6361/201015264}
\bibAnnoteFile{iwasawa2011}

\bibitem[{{Iwasawa} et~al.(2018){Iwasawa}, {U}, {Mazzarella}, {Medling},
  {Sanders}, and {Evans}}]{iwasawa2018}
{Iwasawa}, K., {U}, V., {Mazzarella}, J.~M., {Medling}, A.~M., {Sanders},
  D.~B., and {Evans}, A.~S. (2018).
\newblock {Testing a double AGN hypothesis for Mrk 273}.
\newblock \emph{\aap} 611, A71.
\newblock \doi{10.1051/0004-6361/201731662}
\bibAnnoteFile{iwasawa2018}

\bibitem[{{Jansen} et~al.(2001){Jansen}, {Lumb}, {Altieri}, {Clavel}, {Ehle},
  {Erd} et~al.}]{Jansen2001}
{Jansen}, F., {Lumb}, D., {Altieri}, B., {Clavel}, J., {Ehle}, M., {Erd}, C.,
  et~al. (2001).
\newblock {XMM-Newton observatory. I. The spacecraft and operations}.
\newblock \emph{\aap} 365, L1--L6.
\newblock \doi{10.1051/0004-6361:20000036}
\bibAnnoteFile{Jansen2001}

\bibitem[{{Kammoun} et~al.(2022){Kammoun}, {Barret}, {Peille}, {Willingale},
  {Dauser}, {Wilms} et~al.}]{Kammoun2022}
{Kammoun}, E.~S., {Barret}, D., {Peille}, P., {Willingale}, R., {Dauser}, T.,
  {Wilms}, J., et~al. (2022).
\newblock {The defocused observations of bright sources with Athena/X-IFU}.
\newblock \emph{\aap} 664, A29.
\newblock \doi{10.1051/0004-6361/202243606}
\bibAnnoteFile{Kammoun2022}

\bibitem[{{Komossa} et~al.(2003){Komossa}, {Burwitz}, {Hasinger}, {Predehl},
  {Kaastra}, and {Ikebe}}]{komossa2003}
{Komossa}, S., {Burwitz}, V., {Hasinger}, G., {Predehl}, P., {Kaastra}, J.~S.,
  and {Ikebe}, Y. (2003).
\newblock {Discovery of a Binary Active Galactic Nucleus in the Ultraluminous
  Infrared Galaxy NGC 6240 Using Chandra}.
\newblock \emph{\apjl} 582, L15--L19.
\newblock \doi{10.1086/346145}
\bibAnnoteFile{komossa2003}

\bibitem[{{Kosec} et~al.(2017){Kosec}, {Brightman}, {Stern},
  {M{\"u}ller-S{\'a}nchez}, {Koss}, {Oh} et~al.}]{kosec2017}
{Kosec}, P., {Brightman}, M., {Stern}, D., {M{\"u}ller-S{\'a}nchez}, F.,
  {Koss}, M., {Oh}, K., et~al. (2017).
\newblock {Investigating the Evolution of the Dual AGN System ESO 509-IG066}.
\newblock \emph{\apj} 850, 168.
\newblock \doi{10.3847/1538-4357/aa932e}
\bibAnnoteFile{kosec2017}

\bibitem[{{Koss} et~al.(2012){Koss}, {Mushotzky}, {Treister}, {Veilleux},
  {Vasudevan}, and {Trippe}}]{koss2012}
{Koss}, M., {Mushotzky}, R., {Treister}, E., {Veilleux}, S., {Vasudevan}, R.,
  and {Trippe}, M. (2012).
\newblock {Understanding Dual Active Galactic Nucleus Activation in the nearby
  Universe}.
\newblock \emph{\apjl} 746, L22.
\newblock \doi{10.1088/2041-8205/746/2/L22}
\bibAnnoteFile{koss2012}

\bibitem[{{Koss} et~al.(2016){Koss}, {Glidden}, {Balokovi{\'c}}, {Stern},
  {Lamperti}, {Assef} et~al.}]{koss2016}
{Koss}, M.~J., {Glidden}, A., {Balokovi{\'c}}, M., {Stern}, D., {Lamperti}, I.,
  {Assef}, R., et~al. (2016).
\newblock {NuSTAR Resolves the First Dual AGN above 10 keV in SWIFT
  J2028.5+2543}.
\newblock \emph{\apjl} 824, L4.
\newblock \doi{10.3847/2041-8205/824/1/L4}
\bibAnnoteFile{koss2016}

\bibitem[{{Koss} et~al.(2023){Koss}, {Treister}, {Kakkad}, {Casey-Clyde},
  {Kawamuro}, {Williams} et~al.}]{koss2023}
{Koss}, M.~J., {Treister}, E., {Kakkad}, D., {Casey-Clyde}, J.~A., {Kawamuro},
  T., {Williams}, J., et~al. (2023).
\newblock {UGC 4211: A Confirmed Dual Active Galactic Nucleus in the Local
  Universe at 230 pc Nuclear Separation}.
\newblock \emph{\apjl} 942, L24.
\newblock \doi{10.3847/2041-8213/aca8f0}
\bibAnnoteFile{koss2023}

\bibitem[{{LaMassa} et~al.(2019){LaMassa}, {Yaqoob}, {Boorman}, {Tzanavaris},
  {Levenson}, {Gandhi} et~al.}]{lamassa2019}
{LaMassa}, S.~M., {Yaqoob}, T., {Boorman}, P.~G., {Tzanavaris}, P., {Levenson},
  N.~A., {Gandhi}, P., et~al. (2019).
\newblock {NuSTAR Uncovers an Extremely Local Compton-thick AGN in NGC 4968}.
\newblock \emph{\apj} 887, 173.
\newblock \doi{10.3847/1538-4357/ab552c}
\bibAnnoteFile{lamassa2019}

\bibitem[{{LaMassa} et~al.(2017){LaMassa}, {Yaqoob}, {Levenson}, {Boorman},
  {Heckman}, {Gandhi} et~al.}]{lamassa2017}
{LaMassa}, S.~M., {Yaqoob}, T., {Levenson}, N.~A., {Boorman}, P., {Heckman},
  T.~M., {Gandhi}, P., et~al. (2017).
\newblock {Chandra Reveals Heavy Obscuration and Circumnuclear Star Formation
  in Seyfert 2 Galaxy NGC 4968}.
\newblock \emph{\apj} 835, 91.
\newblock \doi{10.3847/1538-4357/835/1/91}
\bibAnnoteFile{lamassa2017}

\bibitem[{{Laureijs} et~al.(2011){Laureijs}, {Amiaux}, {Arduini},
  {Augu{\`e}res}, {Brinchmann}, {Cole} et~al.}]{laureijs2011}
{Laureijs}, R., {Amiaux}, J., {Arduini}, S., {Augu{\`e}res}, J.~L.,
  {Brinchmann}, J., {Cole}, R., et~al. (2011).
\newblock {Euclid Definition Study Report}.
\newblock \emph{arXiv e-prints} , arXiv:1110.3193\doi{10.48550/arXiv.1110.3193}
\bibAnnoteFile{laureijs2011}

\bibitem[{{Li} et~al.(2021){Li}, {Ballantyne}, and {Bogdanovi{\'c}}}]{li2021}
{Li}, K., {Ballantyne}, D.~R., and {Bogdanovi{\'c}}, T. (2021).
\newblock {The Detectability of Kiloparsec-scale Dual Active Galactic Nuclei:
  The Impact of Galactic Structure and Black Hole Orbital Properties}.
\newblock \emph{\apj} 916, 110.
\newblock \doi{10.3847/1538-4357/ac06a0}
\bibAnnoteFile{li2021}

\bibitem[{{Liu} et~al.(2013){Liu}, {Civano}, {Shen}, {Green}, {Greene}, and
  {Strauss}}]{liu2013}
{Liu}, X., {Civano}, F., {Shen}, Y., {Green}, P., {Greene}, J.~E., and
  {Strauss}, M.~A. (2013).
\newblock {Chandra X-Ray and Hubble Space Telescope Imaging of Optically
  Selected Kiloparsec-scale Binary Active Galactic Nuclei. I. Nature of the
  Nuclear Ionizing Sources}.
\newblock \emph{\apj} 762, 110.
\newblock \doi{10.1088/0004-637X/762/2/110}
\bibAnnoteFile{liu2013}

\bibitem[{{Liu} et~al.(2010){Liu}, {Shen}, {Strauss}, and {Greene}}]{liu2010}
{Liu}, X., {Shen}, Y., {Strauss}, M.~A., and {Greene}, J.~E. (2010).
\newblock {Type 2 Active Galactic Nuclei with Double-Peaked [O III] Lines:
  Narrow-Line Region Kinematics or Merging Supermassive Black Hole Pairs?}
\newblock \emph{\apj} 708, 427--434.
\newblock \doi{10.1088/0004-637X/708/1/427}
\bibAnnoteFile{liu2010}

\bibitem[{{Liu} et~al.(2011){Liu}, {Shen}, {Strauss}, and {Hao}}]{liu2011}
{Liu}, X., {Shen}, Y., {Strauss}, M.~A., and {Hao}, L. (2011).
\newblock {Active Galactic Nucleus Pairs from the Sloan Digital Sky Survey. I.
  The Frequency on \raisebox{-0.5ex}\textasciitilde5-100 kpc Scales}.
\newblock \emph{\apj} 737, 101.
\newblock \doi{10.1088/0004-637X/737/2/101}
\bibAnnoteFile{liu2011}

\bibitem[{{Mannucci} et~al.(2022){Mannucci}, {Pancino}, {Belfiore}, {Cicone},
  {Ciurlo}, {Cresci} et~al.}]{mannucci2022}
{Mannucci}, F., {Pancino}, E., {Belfiore}, F., {Cicone}, C., {Ciurlo}, A.,
  {Cresci}, G., et~al. (2022).
\newblock {Unveiling the population of dual and lensed active galactic nuclei
  at sub-arcsec separations}.
\newblock \emph{Nature Astronomy} 6, 1185--1192.
\newblock \doi{10.1038/s41550-022-01761-5}
\bibAnnoteFile{mannucci2022}

\bibitem[{{Marchesi} et~al.(2018){Marchesi}, {Ajello}, {Marcotulli},
  {Comastri}, {Lanzuisi}, and {Vignali}}]{marchesi2018}
{Marchesi}, S., {Ajello}, M., {Marcotulli}, L., {Comastri}, A., {Lanzuisi}, G.,
  and {Vignali}, C. (2018).
\newblock {Compton-thick AGNs in the NuSTAR Era}.
\newblock \emph{\apj} 854, 49.
\newblock \doi{10.3847/1538-4357/aaa410}
\bibAnnoteFile{marchesi2018}

\bibitem[{{Marchesi} et~al.(2019){Marchesi}, {Ajello}, {Zhao}, {Comastri}, {La
  Parola}, and {Segreto}}]{marchesi2019}
{Marchesi}, S., {Ajello}, M., {Zhao}, X., {Comastri}, A., {La Parola}, V., and
  {Segreto}, A. (2019).
\newblock {Compton-thick AGNs in the NuSTAR Era. V. Joint NuSTAR and XMM-Newton
  Spectral Analysis of Three {\textquotedblleft}Soft-gamma{\textquotedblright}
  Candidate CT-AGNs in the Swift/BAT 100-month Catalog}.
\newblock \emph{\apj} 882, 162.
\newblock \doi{10.3847/1538-4357/ab340a}
\bibAnnoteFile{marchesi2019}

\bibitem[{{Mazzarella} et~al.(2012){Mazzarella}, {Iwasawa}, {Vavilkin},
  {Armus}, {Kim}, {Bothun} et~al.}]{mazzarella2012}
{Mazzarella}, J.~M., {Iwasawa}, K., {Vavilkin}, T., {Armus}, L., {Kim}, D.~C.,
  {Bothun}, G., et~al. (2012).
\newblock {Investigation of Dual Active Nuclei, Outflows, Shock-heated Gas, and
  Young Star Clusters in Markarian 266}.
\newblock \emph{\aj} 144, 125.
\newblock \doi{10.1088/0004-6256/144/5/125}
\bibAnnoteFile{mazzarella2012}

\bibitem[{{Meidinger} et~al.(2020){Meidinger}, {Albrecht}, {Beitler},
  {Bonholzer}, {Emberger}, {Frank} et~al.}]{meidinger2020}
{Meidinger}, N., {Albrecht}, S., {Beitler}, C., {Bonholzer}, M., {Emberger},
  V., {Frank}, J., et~al. (2020).
\newblock {Development status of the wide field imager instrument for Athena}.
\newblock In \emph{Society of Photo-Optical Instrumentation Engineers (SPIE)
  Conference Series}. vol. 11444 of \emph{Society of Photo-Optical
  Instrumentation Engineers (SPIE) Conference Series}, 114440T.
\newblock \doi{10.1117/12.2560507}
\bibAnnoteFile{meidinger2020}

\bibitem[{{M{\"u}ller-S{\'a}nchez} et~al.(2015){M{\"u}ller-S{\'a}nchez},
  {Comerford}, {Nevin}, {Barrows}, {Cooper}, and {Greene}}]{mullersanchez2015}
{M{\"u}ller-S{\'a}nchez}, F., {Comerford}, J.~M., {Nevin}, R., {Barrows},
  R.~S., {Cooper}, M.~C., and {Greene}, J.~E. (2015).
\newblock {The Origin of Double-peaked Narrow Lines in Active Galactic Nuclei.
  I. Very Large Array Detections of Dual AGNs and AGN Outflows}.
\newblock \emph{\apj} 813, 103.
\newblock \doi{10.1088/0004-637X/813/2/103}
\bibAnnoteFile{mullersanchez2015}

\bibitem[{{M{\"u}ller-S{\'a}nchez} et~al.(2018){M{\"u}ller-S{\'a}nchez},
  {Nevin}, {Comerford}, {Davies}, {Privon}, and
  {Treister}}]{muller-sanchez2018}
{M{\"u}ller-S{\'a}nchez}, F., {Nevin}, R., {Comerford}, J.~M., {Davies}, R.~I.,
  {Privon}, G.~C., and {Treister}, E. (2018).
\newblock {Two separate outflows in the dual supermassive black hole system NGC
  6240}.
\newblock \emph{\nat} 556, 345--348.
\newblock \doi{10.1038/s41586-018-0033-2}
\bibAnnoteFile{muller-sanchez2018}

\bibitem[{{Nandra} et~al.(2013){Nandra}, {Barret}, {Barcons}, {Fabian}, {den
  Herder}, {Piro} et~al.}]{Nandra2013}
{Nandra}, K., {Barret}, D., {Barcons}, X., {Fabian}, A., {den Herder}, J.-W.,
  {Piro}, L., et~al. (2013).
\newblock {The Hot and Energetic Universe: A White Paper presenting the science
  theme motivating the Athena+ mission}.
\newblock \emph{arXiv e-prints} , arXiv:1306.2307\doi{10.48550/arXiv.1306.2307}
\bibAnnoteFile{Nandra2013}

\bibitem[{{Nardini}(2017)}]{nardini2017}
{Nardini}, E. (2017).
\newblock {Nuclear absorption and emission in the AGN merger NGC 6240 : the
  hard X-ray view}.
\newblock \emph{\mnras} 471, 3483--3493.
\newblock \doi{10.1093/mnras/stx1878}
\bibAnnoteFile{nardini2017}

\bibitem[{{Oda} et~al.(2018){Oda}, {Ueda}, {Tanimoto}, and {Ricci}}]{oda2018}
{Oda}, S., {Ueda}, Y., {Tanimoto}, A., and {Ricci}, C. (2018).
\newblock {Hard X-Ray View of HCG 16 (Arp 318)}.
\newblock \emph{\apj} 855, 79.
\newblock \doi{10.3847/1538-4357/aaaccc}
\bibAnnoteFile{oda2018}

\bibitem[{{P{\'e}rez-Torres} et~al.(2010){P{\'e}rez-Torres}, {Alberdi},
  {Romero-Ca{\~n}izales}, and {Bondi}}]{pereztorres2010}
{P{\'e}rez-Torres}, M.~A., {Alberdi}, A., {Romero-Ca{\~n}izales}, C., and
  {Bondi}, M. (2010).
\newblock {Serendipitous discovery of the long-sought active galactic nucleus
  in Arp 299-A}.
\newblock \emph{\aap} 519, L5.
\newblock \doi{10.1051/0004-6361/201015462}
\bibAnnoteFile{pereztorres2010}

\bibitem[{{Pfeifle} et~al.(2019{\natexlab{a}}){Pfeifle}, {Satyapal},
  {Manzano-King}, {Cann}, {Sexton}, {Rothberg} et~al.}]{pfeifle2019b}
{Pfeifle}, R.~W., {Satyapal}, S., {Manzano-King}, C., {Cann}, J., {Sexton},
  R.~O., {Rothberg}, B., et~al. (2019{\natexlab{a}}).
\newblock {A Triple AGN in a Mid-infrared Selected Late-stage Galaxy Merger}.
\newblock \emph{\apj} 883, 167.
\newblock \doi{10.3847/1538-4357/ab3a9b}
\bibAnnoteFile{pfeifle2019b}

\bibitem[{{Pfeifle} et~al.(2019{\natexlab{b}}){Pfeifle}, {Satyapal}, {Secrest},
  {Gliozzi}, {Ricci}, {Ellison} et~al.}]{pfeifle2019a}
{Pfeifle}, R.~W., {Satyapal}, S., {Secrest}, N.~J., {Gliozzi}, M., {Ricci}, C.,
  {Ellison}, S.~L., et~al. (2019{\natexlab{b}}).
\newblock {Buried Black Hole Growth in IR-selected Mergers: New Results from
  Chandra}.
\newblock \emph{\apj} 875, 117.
\newblock \doi{10.3847/1538-4357/ab07bc}
\bibAnnoteFile{pfeifle2019a}

\bibitem[{{Pfeifle} et~al.(2023){Pfeifle}, {Weaver}, {Satyapal}, {Ricci},
  {Secrest}, {Gliozzi} et~al.}]{pfeifle2023c}
{Pfeifle}, R.~W., {Weaver}, K., {Satyapal}, S., {Ricci}, C., {Secrest}, N.~J.,
  {Gliozzi}, M., et~al. (2023).
\newblock {NuSTAR Observations of Four Mid-IR Selected Dual AGN Candidates in
  Galaxy Mergers}.
\newblock \emph{arXiv e-prints} ,
  arXiv:2306.16437\doi{10.48550/arXiv.2306.16437}
\bibAnnoteFile{pfeifle2023c}

\bibitem[{{Piconcelli} et~al.(2010){Piconcelli}, {Vignali}, {Bianchi},
  {Mathur}, {Fiore}, {Guainazzi} et~al.}]{piconcelli2010}
{Piconcelli}, E., {Vignali}, C., {Bianchi}, S., {Mathur}, S., {Fiore}, F.,
  {Guainazzi}, M., et~al. (2010).
\newblock {Witnessing the Key Early Phase of Quasar Evolution: An Obscured
  Active Galactic Nucleus Pair in the Interacting Galaxy IRAS 20210+1121}.
\newblock \emph{\apjl} 722, L147--L151.
\newblock \doi{10.1088/2041-8205/722/2/L147}
\bibAnnoteFile{piconcelli2010}

\bibitem[{{Ptak} et~al.(2015){Ptak}, {Hornschemeier}, {Zezas}, {Lehmer},
  {Yukita}, {Wik} et~al.}]{ptak2015}
{Ptak}, A., {Hornschemeier}, A., {Zezas}, A., {Lehmer}, B., {Yukita}, M.,
  {Wik}, D., et~al. (2015).
\newblock {A Focused, Hard X-Ray Look at Arp 299 with NuSTAR}.
\newblock \emph{\apj} 800, 104.
\newblock \doi{10.1088/0004-637X/800/2/104}
\bibAnnoteFile{ptak2015}

\bibitem[{{Ricci} et~al.(2017){Ricci}, {Bauer}, {Treister}, {Schawinski},
  {Privon}, {Blecha} et~al.}]{ricci2017MNRAS}
{Ricci}, C., {Bauer}, F.~E., {Treister}, E., {Schawinski}, K., {Privon}, G.~C.,
  {Blecha}, L., et~al. (2017).
\newblock {Growing supermassive black holes in the late stages of galaxy
  mergers are heavily obscured}.
\newblock \emph{\mnras} 468, 1273--1299.
\newblock \doi{10.1093/mnras/stx173}
\bibAnnoteFile{ricci2017MNRAS}

\bibitem[{{Ricci} et~al.(2023){Ricci}, {Chang}, {Kawamuro}, {Privon},
  {Mushotzky}, {Trakhtenbrot} et~al.}]{ricci2023}
{Ricci}, C., {Chang}, C.-S., {Kawamuro}, T., {Privon}, G.~C., {Mushotzky}, R.,
  {Trakhtenbrot}, B., et~al. (2023).
\newblock {A Tight Correlation between Millimeter and X-Ray Emission in
  Accreting Massive Black Holes from <100 mas Resolution ALMA Observations}.
\newblock \emph{\apjl} 952, L28.
\newblock \doi{10.3847/2041-8213/acda27}
\bibAnnoteFile{ricci2023}

\bibitem[{{Ricci} et~al.(2021){Ricci}, {Privon}, {Pfeifle}, {Armus}, {Iwasawa},
  {Torres-Alb{\`a}} et~al.}]{ricci2021}
{Ricci}, C., {Privon}, G.~C., {Pfeifle}, R.~W., {Armus}, L., {Iwasawa}, K.,
  {Torres-Alb{\`a}}, N., et~al. (2021).
\newblock {A hard X-ray view of luminous and ultra-luminous infrared galaxies
  in GOALS - I. AGN obscuration along the merger sequence}.
\newblock \emph{\mnras} 506, 5935--5950.
\newblock \doi{10.1093/mnras/stab2052}
\bibAnnoteFile{ricci2021}

\bibitem[{{Rubinur} et~al.(2019){Rubinur}, {Das}, and {Kharb}}]{rubinur2019}
{Rubinur}, K., {Das}, M., and {Kharb}, P. (2019).
\newblock {Searching for dual AGN in galaxies with double-peaked emission line
  spectra using radio observations}.
\newblock \emph{\mnras} 484, 4933--4950.
\newblock \doi{10.1093/mnras/stz334}
\bibAnnoteFile{rubinur2019}

\bibitem[{{Satyapal} et~al.(2014){Satyapal}, {Ellison}, {McAlpine}, {Hickox},
  {Patton}, and {Mendel}}]{satyapal2014}
{Satyapal}, S., {Ellison}, S.~L., {McAlpine}, W., {Hickox}, R.~C., {Patton},
  D.~R., and {Mendel}, J.~T. (2014).
\newblock {Galaxy pairs in the Sloan Digital Sky Survey - IX. Merger-induced
  AGN activity as traced by the Wide-field Infrared Survey Explorer}.
\newblock \emph{\mnras} 441, 1297--1304.
\newblock \doi{10.1093/mnras/stu650}
\bibAnnoteFile{satyapal2014}

\bibitem[{{Satyapal} et~al.(2017){Satyapal}, {Secrest}, {Ricci}, {Ellison},
  {Rothberg}, {Blecha} et~al.}]{satyapal2017}
{Satyapal}, S., {Secrest}, N.~J., {Ricci}, C., {Ellison}, S.~L., {Rothberg},
  B., {Blecha}, L., et~al. (2017).
\newblock {Buried AGNs in Advanced Mergers: Mid-infrared Color Selection as a
  Dual AGN Candidate Finder}.
\newblock \emph{\apj} 848, 126.
\newblock \doi{10.3847/1538-4357/aa88ca}
\bibAnnoteFile{satyapal2017}

\bibitem[{{Schawinski} et~al.(2015){Schawinski}, {Koss}, {Berney}, and
  {Sartori}}]{schawinski2015}
{Schawinski}, K., {Koss}, M., {Berney}, S., and {Sartori}, L.~F. (2015).
\newblock {Active galactic nuclei flicker: an observational estimate of the
  duration of black hole growth phases of {\ensuremath{\sim}}{}10$^{5}$ yr}.
\newblock \emph{\mnras} 451, 2517--2523.
\newblock \doi{10.1093/mnras/stv1136}
\bibAnnoteFile{schawinski2015}

\bibitem[{{Schwartzman} et~al.(2023){Schwartzman}, {Clarke}, {Nyland},
  {Secrest}, {Pfeifle}, {Schmitt} et~al.}]{schwartzman2023}
{Schwartzman}, E., {Clarke}, T.~E., {Nyland}, K., {Secrest}, N.~J., {Pfeifle},
  R.~W., {Schmitt}, H.~R., et~al. (2023).
\newblock {VaDAR: Varstrometry for Dual AGN using Radio interferometry}.
\newblock \emph{arXiv e-prints} ,
  arXiv:2306.13219\doi{10.48550/arXiv.2306.13219}
\bibAnnoteFile{schwartzman2023}

\bibitem[{{Shen} et~al.(2021){Shen}, {Chen}, {Hwang}, {Liu}, {Zakamska},
  {Oguri} et~al.}]{shen2021}
{Shen}, Y., {Chen}, Y.-C., {Hwang}, H.-C., {Liu}, X., {Zakamska}, N., {Oguri},
  M., et~al. (2021).
\newblock {A hidden population of high-redshift double quasars unveiled by
  astrometry}.
\newblock \emph{Nature Astronomy} 5, 569--574.
\newblock \doi{10.1038/s41550-021-01323-1}
\bibAnnoteFile{shen2021}

\bibitem[{{Shen} et~al.(2019){Shen}, {Hwang}, {Zakamska}, and {Liu}}]{shen2019}
{Shen}, Y., {Hwang}, H.-C., {Zakamska}, N., and {Liu}, X. (2019).
\newblock {Varstrometry for Off-nucleus and Dual Sub-Kpc AGN (VODKA): How Well
  Centered Are Low-z AGN?}
\newblock \emph{\apjl} 885, L4.
\newblock \doi{10.3847/2041-8213/ab4b54}
\bibAnnoteFile{shen2019}

\bibitem[{Skilling(2004)}]{Skilling2004}
Skilling, J. (2004).
\newblock Nested sampling.
\newblock \emph{AIP Conference Proceedings} 735, 395.
\newblock \doi{10.1063/1.1835238}
\bibAnnoteFile{Skilling2004}

\bibitem[{{Smith} et~al.(2010){Smith}, {Shields}, {Bonning}, {McMullen},
  {Rosario}, and {Salviander}}]{smith2010}
{Smith}, K.~L., {Shields}, G.~A., {Bonning}, E.~W., {McMullen}, C.~C.,
  {Rosario}, D.~J., and {Salviander}, S. (2010).
\newblock {A Search for Binary Active Galactic Nuclei: Double-peaked [O III]
  AGNs in the Sloan Digital Sky Survey}.
\newblock \emph{\apj} 716, 866--877.
\newblock \doi{10.1088/0004-637X/716/1/866}
\bibAnnoteFile{smith2010}

\bibitem[{{Toomre} and {Toomre}(1972)}]{toomre1972}
{Toomre}, A. and {Toomre}, J. (1972).
\newblock {Galactic Bridges and Tails}.
\newblock \emph{\apj} 178, 623--666.
\newblock \doi{10.1086/151823}
\bibAnnoteFile{toomre1972}

\bibitem[{{Torres-Alb{\`a}} et~al.(2018){Torres-Alb{\`a}}, {Iwasawa},
  {D{\'\i}az-Santos}, {Charmandaris}, {Ricci}, {Chu} et~al.}]{torres-alba2018}
{Torres-Alb{\`a}}, N., {Iwasawa}, K., {D{\'\i}az-Santos}, T., {Charmandaris},
  V., {Ricci}, C., {Chu}, J.~K., et~al. (2018).
\newblock {C-GOALS. II. Chandra observations of the lower luminosity sample of
  nearby luminous infrared galaxies in GOALS}.
\newblock \emph{\aap} 620, A140.
\newblock \doi{10.1051/0004-6361/201834105}
\bibAnnoteFile{torres-alba2018}

\bibitem[{{Van Wassenhove} et~al.(2012){Van Wassenhove}, {Volonteri}, {Mayer},
  {Dotti}, {Bellovary}, and {Callegari}}]{vanwassenhove2012}
{Van Wassenhove}, S., {Volonteri}, M., {Mayer}, L., {Dotti}, M., {Bellovary},
  J., and {Callegari}, S. (2012).
\newblock {Observability of Dual Active Galactic Nuclei in Merging Galaxies}.
\newblock \emph{\apjl} 748, L7.
\newblock \doi{10.1088/2041-8205/748/1/L7}
\bibAnnoteFile{vanwassenhove2012}

\bibitem[{{Wang} et~al.(2009){Wang}, {Chen}, {Hu}, {Mao}, {Zhang}, and
  {Bian}}]{wang2009}
{Wang}, J.-M., {Chen}, Y.-M., {Hu}, C., {Mao}, W.-M., {Zhang}, S., and {Bian},
  W.-H. (2009).
\newblock {Active Galactic Nuclei with Double-Peaked Narrow Lines: Are they
  Dual Active Galactic Nuclei?}
\newblock \emph{\apjl} 705, L76--L80.
\newblock \doi{10.1088/0004-637X/705/1/L76}
\bibAnnoteFile{wang2009}

\bibitem[{{Weston} et~al.(2017){Weston}, {McIntosh}, {Brodwin}, {Mann},
  {Cooper}, {McConnell} et~al.}]{weston2017}
{Weston}, M.~E., {McIntosh}, D.~H., {Brodwin}, M., {Mann}, J., {Cooper}, A.,
  {McConnell}, A., et~al. (2017).
\newblock {Incidence of WISE -selected obscured AGNs in major mergers and
  interactions from the SDSS}.
\newblock \emph{\mnras} 464, 3882--3906.
\newblock \doi{10.1093/mnras/stw2620}
\bibAnnoteFile{weston2017}

\bibitem[{{Yamada} et~al.(2018){Yamada}, {Ueda}, {Oda}, {Tanimoto}, {Imanishi},
  {Terashima} et~al.}]{yamada2018}
{Yamada}, S., {Ueda}, Y., {Oda}, S., {Tanimoto}, A., {Imanishi}, M.,
  {Terashima}, Y., et~al. (2018).
\newblock {Broadband X-Ray Spectral Analysis of the Double-nucleus Luminous
  Infrared Galaxy Mrk 463}.
\newblock \emph{\apj} 858, 106.
\newblock \doi{10.3847/1538-4357/aabacb}
\bibAnnoteFile{yamada2018}

\bibitem[{{Zhou} et~al.(2004){Zhou}, {Wang}, {Zhang}, {Dong}, and
  {Li}}]{zhou2004}
{Zhou}, H., {Wang}, T., {Zhang}, X., {Dong}, X., and {Li}, C. (2004).
\newblock {Obscured Binary Quasar Cores in SDSS J104807.74+005543.5?}
\newblock \emph{\apjl} 604, L33--L36.
\newblock \doi{10.1086/383310}
\bibAnnoteFile{zhou2004}

\bibitem[{{ZuHone} et~al.(2023){ZuHone}, {Vikhlinin}, {Tremblay}, {Randall},
  {Andrade-Santos}, and {Bourdin}}]{zuhone2023}
[Dataset] {ZuHone}, J.~A., {Vikhlinin}, A., {Tremblay}, G.~R., {Randall},
  S.~W., {Andrade-Santos}, F., and {Bourdin}, H. (2023).
\newblock {SOXS: Simulated Observations of X-ray Sources}.
\newblock Astrophysics Source Code Library, record ascl:2301.024
\bibAnnoteFile{zuhone2023}

\end{thebibliography}

%%% Make sure to upload the bib file along with the tex file and PDF
%%% Please see the test.bib file for some examples of references

% \section*{Figure captions}

%%% Please be aware that for original research articles we only permit a combined number of 15 figures and tables, one figure with multiple subfigures will count as only one figure.
%%% Use this if adding the figures directly in the mansucript, if so, please remember to also upload the files when submitting your article
%%% There is no need for adding the file termination, as long as you indicate where the file is saved. In the examples below the files (logo1.eps and logos.eps) are in the Frontiers LaTeX folder
%%% If using *.tif files convert them to .jpg or .png
%%%  NB logo1.eps is required in the path in order to correctly compile front page header %%%

\end{document}